\documentclass[11pt]{article}%
\usepackage[onehalfspacing]{setspace}
\usepackage[left=1.1in,right=1.1in,top=1.1in,bottom=1.1in]{geometry}
\usepackage[colorlinks=true,linkcolor=RawSienna,citecolor=RawSienna,urlcolor=RawSienna,bookmarksopen=true,pdfstartview=FitB]%
{hyperref}
\usepackage[dvipsnames]{xcolor}
\usepackage{natbib}
\usepackage{amsmath}
\usepackage{amsfonts}
\usepackage{amssymb}
\usepackage{graphicx}
\usepackage{float}
\usepackage{tikz}
\usepackage{hyphenat}%
\providecommand{\U}[1]{\protect\rule{.1in}{.1in}}

\usetikzlibrary{shapes.geometric, shapes, arrows}
\usetikzlibrary{arrows.meta}
\usetikzlibrary{fit,calc,positioning}
\usetikzlibrary{trees,arrows,automata,graphs}
\usetikzlibrary{positioning,chains,fit,shapes,calc,topaths}
\usetikzlibrary{decorations.markings}
\newtheorem{theorem}{Theorem}

\newtheorem{definition}{Definition}
\newtheorem{lemma}{Lemma}
\newtheorem{proposition}{Proposition}

\allowdisplaybreaks
\begin{document}

\title{\vspace{-1.2cm}A grouped, selectively weighted false discovery rate procedure}
\author{Xiongzhi Chen\thanks{Corresponding author: Department of Mathematics and
Statistics, Washington State University, Pullman, WA 99164, USA; Email:
\texttt{xiongzhi.chen@wsu.edu}.} \ and Sanat K. Sarkar\thanks{Department of
Statistical Science and Fox School of Business, Temple University,
Philadelphia, PA 19122, USA; Email: \texttt{sanat@temple.edu}.} }
\date{}
\maketitle

\begin{abstract}
False discovery rate (FDR) control in structured hypotheses testing is an
important topic in simultaneous inference. Most existing methods that aim to
utilize group structure among hypotheses either employ the groupwise mixture
model or weight all p-values or hypotheses. Thus, their powers can be improved
when the groupwise mixture model is inappropriate or when most groups contain
only true null hypotheses. Motivated by this, we propose a grouped,
selectively weighted FDR procedure, which we refer to as ``sGBH''. Specifically, without employing the
groupwise mixture model, sGBH identifies groups of hypotheses of interest,
weights p-values in each such group only, and tests only the selected hypotheses using the weighted p-values.
The sGBH subsumes a standard grouped, weighted FDR procedure which we refer to as ``GBH''.
We provide simple conditions to ensure the conservativeness of sGBH, together with empirical evidence on its much improved power over GBH.
The new procedure is applied to a gene expression study.

\end{abstract}

\section{Introduction}

\label{secIntro}

Multiple testing aiming at false discovery rate (FDR) control has been
routinely applied in genomics, genetics, neuroimaging, drug safety study and
other fields. In many multiple testing scenarios, there is prior information
on certain characteristics of hypotheses or statistics. For example, groups of hypotheses may have
different proportions of true nulls, or statistics associated with a group of
hypotheses may possess the same type of dependency structure or have similar
powers. To utilize such information, methods based on hypotheses grouping and
weighting \citep{Liu:2016,Basu:2017} or p-value grouping and weighting
\citep{Cai:2009,Hu:2010,Chen:2015discretefdr,Nandi:2018} have been developed.

Even though these methods can often be more powerful than the procedures of
\cite{Benjamini:1995}, \cite{Benjamini:2000} and \cite{Sun:2007}, they have
some limitations. For example, those in the Bayesian paradigm, e.g.,
``TLTA'' in \cite{Liu:2016}, employ the
groupwise mixture model where component densities are assumed to be
continuous, whereas those in the frequentist paradigm, e.g., the ``GBH''
procedures of \cite{Hu:2010,Nandi:2018}, weight each p-value by treating each group of
hypotheses equally importantly.

In contrast, there are many multiple testing settings with the following two
features: (i) the number of false nulls is relatively small and some groups of
hypotheses may contain no false nulls, which happens in, e.g., differential
gene expression studies based on microarrays, and (ii) the null distributions
of test statistics or p-values are different from each other and the groupwise
mixture model is inappropriate, as in, e.g., differential gene expression
studies based on discrete RNA-seq data. Even though \textquotedblleft
TLTA\textquotedblright\ can accommodate (i) by taking into account false
discoveries both within and between groups and GBH accommodates (ii) for
sub-uniform p-values with heterogeneous null distributions, neither of them is
able to accommodate both (i) and (ii).

\subsection{Main contribution}

\begin{figure}[t]
\centering
\includegraphics[height=0.09\textheight, width=0.95\textwidth]{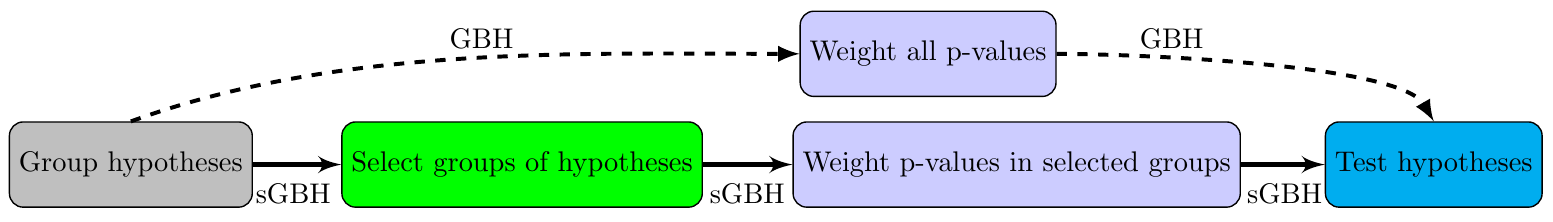}\caption[iGBH]%
{A schematic comparison between GBH and sGBH. sGBH selects groups of
hypotheses of interest and weights only p-values in each such group, whereas
GBH does not select groups and weights p-values in each group.}%
\label{fig:igbh}%
\end{figure}

We propose \textquotedblleft sGBH\textquotedblright, a grouped and selectively
weighted multiple testing procedure as a refinement and extension of GBH. A
schematic comparison between GBH and sGBH is given in \autoref{fig:igbh}, and
the details on sGBH are provided in \autoref{secOracleiGBH}. Unlike GBH,
sGBH first identifies groups of hypotheses of interest and then weights p-values in each
such group only. It reduces to GBH when all groups are of interest, and
whenever this is the case, results that hold for GBH hold for sGBH also.
Compared to TLTA, sGBH does not depend on the groupwise mixture
model, and its weights are either induced by estimators of the groupwise
proportions or those introduced by \cite{Nandi:2018}. In essence, sGBH
integrates the appealing features of both GBH and TLTA, and is able to account
for each of the two features mentioned earlier.

When adapting sGBH to data through estimating the oracle weights, we consider two different ways of doing it and thus produce two adaptive versions of sGBH.
In version one, the weights are induced by the estimated groupwise proportions of true nulls, whereas in
version two, the weights are borrowed from \cite{Nandi:2018}. Unlike
\cite{Hu:2010}, who assumed the convergence of various empirical processes
related to p-values and the asymptotic conservativeness of each estimator of
the groupwise proportion of true nulls, we provide simple conditions on the
conservativeness of the first version of the adaptive sGBH. The key to achieve
this is a reciprocally conservative or consistent estimator of the proportion
of true nulls. As such, our method presents a general strategy to
non-asymptotically bound the FDR and show the conservativeness of an
adaptively weighted and grouped testing procedure whose weights are induced by
the estimated groupwise proportions of true nulls. On the other hand, once
groups with false nulls are selected and groups with all true nulls
are correctly identified but not selected, the second version of the adaptive sGBH
reduces to the one-way GBH procedure of \cite{Nandi:2018} and is automatically
conservative non-asymptotically under independence.

In addition, we propose a variant of sGBH and justify its conservativeness
under similar conditions to those for the first version of the adaptive sGBH.
We show that under a ``sparse configuration'' where each interesting group of hypotheses contains
some false nulls and the rest all true nulls, neither the oracle sGBH nor the
variant can reject more false nulls than the oracle GBH, even though
the oracle GBH cannot be implemented and its power not obtainable in non-asymptotic settings. Specifically,
our simulations show that, under nontrivial sparse configurations, both
versions of the adaptive sGBH are conservative and usually more
powerful than their corresponding versions of the adaptive GBH. Further, we
argue that outside nontrivial sparse configurations, the variant and its
adaptive version respectively can be more powerful than the oracle GBH and its
adaptive version, even though we were not able to theoretically identify
conditions that guarantee so. These findings are perhaps the first on better
adaptation to group structures of hypotheses for FDR control in multiple testing in the frequentist paradigm.

\subsection{Organization of article}

The rest of the article is organized as follows. \autoref{secSettings}
introduces sGBH and investigates its conservativeness, and \autoref{secAdap}
provides two versions of the adaptive sGBH. A simulation study on the adaptive
sGBH is given in \autoref{secSim}, and an application in \autoref{secApp}. The
article ends with a discussion in \autoref{secDis}. Proofs related to sGBH are given in the appendix, and a variant of
sGBH and some simulations results are provided in the supplementary material.

\section{Grouped hypotheses testing and sGBH}

\label{secSettings}

Consider simultaneously testing $m$ null hypotheses $\left\{  H_{i}\right\}
_{i=1}^{m}$ with their corresponding p-values $\left\{  p_{i}\right\}
_{i=1}^{m}$. Unless otherwise noted, each $p_{i}$ is assumed to be
``super-uniform'' under its associated null
hypothesis, i.e., its null distribution $F_{i}$ satisfies $F_{i}\left(
t\right)  \leq t$ for $t\in\left[  0,1\right]  $.
We assume throughout this article that $\min\left\{  p_{i}:i\geq1\right\}  >0$
almost surely, in order to avoid the undetermined operation $0\times\infty$
when a p-value is $0$ and a weight is $\infty$.

Let $\alpha\in\left(0,1\right)  $ be a nominal FDR level.
Recall the BH procedure of \cite{Benjamini:1995} (``BH'' for short) as follows: let $\left\{  p_{\left(  i\right)
}\right\}  _{i=1}^{m}$ be the order statistics (non-decreasing in $i$) of
$\left\{  p_{i}\right\}  _{i=1}^{m}$, and $H_{\left(  i\right)  }$ the null
hypothesis associated with $p_{\left(  i\right)  }$ for each $i$; set
$\theta=\max\left\{  i:p_{\left(  i\right)  }\leq\frac{i}{m}\alpha\right\}  $,
and reject $H_{\left(  j\right)  }$ for $1\leq j\leq\theta$ if $\left\{
i:p_{\left(  i\right)  }\leq\frac{i}{m}\alpha\right\}  \neq\varnothing$ but reject none
otherwise.

The BH is an ungrouped testing procedure. In contrast, the settings for grouped hypotheses testing are described as follows.
For each natural number $s$, let $\mathbb{N}_{s}=\left\{  1,\ldots,s\right\}
$. Let $I_{0}$ be the index set of true nulls with cardinality $m_{0}$,
$\pi_{0}=m_{0}m^{-1}$ be the proportion of true nulls, and $\pi=1-\pi_{0}$ be
the proportion of false nulls. Let the $l$ non-empty sets
$\left\{  {G_{j}}\right\}_{j=1}^{l}$ be a partition of $\mathbb{N}_{m}$, and accordingly let $\left\{  H_{i}\right\}
_{i=1}^{m}$ be partitioned into $\mathcal{H}_{j}=\left\{  H_{j_{k}}:k\in
G_{j}\right\}  $ for $j\in\mathbb{N}_{l}$. For each $j$, let $n_{j}$ be
the cardinality of $G_{j}$, $\pi_{j0}$ be the proportion of true nulls for
$\mathcal{H}_{j}$, and $\pi_{j1}=1-\pi_{j0}$. We refer to the partition and
its associated proportions of true nulls and cardinalities as
``a hypotheses configuration''.
Note that%
\begin{equation}
\pi_{0}=m^{-1}\sum\nolimits_{j=1}^{l}\pi_{j0}n_{j}. \label{eqe1}%
\end{equation}

\subsection{The oracle sGBH and its properties}

\label{secOracleiGBH}

Let $S$ be a subset of $\mathbb{N}_{l}$, and call each $\mathcal{H}_{j}$, $j\in$ $S$ an ``interesting
group'' and $\mathcal{H}_{j}$, $j\notin$ $S$ an ``uninteresting group''.
The oracle sGBH is described as follows: (1) set $p_{j_{k}}=\infty$ for each $j\notin S$ and $k\in
G_{j}$, i.e., accept the hypothesis set $\mathcal{H}_{S^{\prime}}=\left\{H_{j_{k}}:j\notin S,k\in G_{j}\right\}  $; (2)
obtain the proportion of true nulls among all interesting groups as
\begin{equation}
\tilde{\pi}_{0}=\sum\nolimits_{j\in S}\pi_{j0}n_{j}\left(  \sum\nolimits_{j\in
S}n_{j}\right)  ^{-1}, \label{eq10a1}%
\end{equation}
define the weight%
\begin{equation}
v_{j}=\frac{\pi_{j0}\left(  1-\tilde{\pi}_{0}\right)  }{1-\pi_{j0}}\text{ \ for each \ }j\in S, \label{eqe2}%
\end{equation}
and weight the p-value $p_{j_{k}}$ into $\tilde{p}_{j_{k}}=p_{j_{k}}v_{j}$ for each $k\in G_{j}$ and $j\in S$;
(3) when $\tilde{\pi}_{0}=1$, no rejections are made; otherwise, apply BH to the weighted p-values in the interesting groups, i.e., to the p-value set $\tilde{\mathbf{p}}
_{S}=\left\{  \tilde{p}_{j_{k}}:j\in S,k\in G_{j}\right\}  $ and their corresponding nulls.

We have three remarks regarding the oracle sGBH: firstly, practically speaking $S$ contains groups of hypotheses each of which contains some false nulls;
secondly, $v_{j}=\infty$ is set when $\tilde{\pi}_{0}=1$ and/or $\pi_{j0}=1$; thirdly, the oracle GBH of \cite{Hu:2010} always sets $S = \mathbb{N}_{l}$ and weighs all $m$ p-values, and is hence subsumed by the oracle sGBH.

Since in practice a group of hypotheses may or may not contain any false nulls, we introduce

\begin{definition}
A ``sparse configuration'' for
$\left\{  H_{i}\right\}  _{i=1}^{m}$ is such that $\pi_{j0}=1$ for
$\mathcal{H}_{j}$ with $j\notin S$ but $\pi_{j0}<1$ for $\mathcal{H}_{j}$ with
$j\in S$ for a subset $S$ of $\mathbb{N}_{l}$. When $S\neq\varnothing$, a
sparse configuration is called ``nontrivial'';
otherwise, it is called ``trivial''.
\end{definition}
\noindent
A nontrivial sparse configuration excludes the trivial case where
all the $m$ null hypotheses are true, and in most statistical applications a hypotheses configuration is sparse.

Set $\mathbf{p}=\left(p_{1},\ldots,p_{m}\right)  $.
Recall from \cite{Benjamini:2001} the property ``positive regression dependency on each one from
$I_{0}$ (PRDS)'', i.e., for any measurable non-decreasing set
$D\subseteq\left[  0,1\right]  ^{m}$, the function $t\mapsto\Pr\left(  \left.
\mathbf{p}\in D\right\vert p_{i}=t\right)  $ is nondecreasing for each $i\in
I_{0}$.  We are ready to assert the conservativeness of the oracle sGBH under PRDS via the following theorem.

\begin{theorem}
\label{ThmOracle}Consider a sparse configuration for $\left\{  H_{i}\right\}
_{i=1}^{m}$. If $S=\mathbb{N}_{l}$ or $\varnothing$, then the oracle
sGBH coincides with the oracle GBH. However, when $S\neq\varnothing$, the oracle sGBH and the oracle GBH reject
the same set of hypotheses at the same nominal FDR level. Further, when $\left\{  p_{i}\right\}
_{i=1}^{m}$ have the property of PRDS, the oracle sGBH is conservative.

\end{theorem}

Even though \autoref{ThmOracle} asserts that the oracle sGBH and the oracle
GBH coincide under nontrivial sparse configurations, the former procedure not only reduces the
complexity of the latter by applying the BH procedure only to p-values from
the interesting groups but also helps gauge, as we will provide some numerical
evidence, the reduced power of GBH when the set of interesting groups is not
always correctly estimated under a nontrivial sparse configuration.
In \autoref{secQGBH} we will introduce a quasi-adaptive form of sGBH, as a precursor to the (fully) data-adaptive sGBH to be developed in \autoref{secAdap}.

\subsection{The quasi-adaptive sGBH}

\label{secQGBH}

Let the ``quasi-adaptive sGBH (qGBH)'' be
such that $S$ in the oracle sGBH is replaced by its estimate $\hat{S}$ but
each $\pi_{j0}$, $j\in\mathbb{N}_{l}$ is retained. Under a nontrivial sparse
configuration, qGBH interpolates the oracle GBH, the oracle sGBH and their
adaptive versions. In particular, the power difference between the oracle GBH
and qGBH reflects the impact of estimating $S$ on the oracle GBH.

We have an interesting result on the conservativeness of qGBH:

\begin{proposition}
\label{PropQgbh}Consider a nontrivial sparse configuration where $S$ has cardinality $1$. Then the qGBH is conservative when p-values satisfy PRDS
regardless of how $\hat{S}$ is constructed.
\end{proposition}

\noindent\autoref{PropQgbh} also implies that when there is only one
interesting group, the adaptive sGBH (to be introduced in
\autoref{secAdap}) is asymptotically conservative when it employs
consistent estimators of each groupwise proportion of true nulls. In
contrast to the claim of \autoref{PropQgbh}, when the cardinality of $S$ is
greater than $1$ under a nontrivial sparse configuration, the involved
probability estimates are rather complicated and somewhat intractable, and it
is much harder to derive a concise, relatively tight upper bound on the
FDR\ of qGBH.

A simulation study involving Gaussian distributional setting has been carried out to compare the qGBH with the oracle
GBH under the same simulation design provided in \autoref{simDesign}, except
that the minimal nonzero Normal means has magnitude $\mu_{\ast}=0.1$. For this
simulation, $\hat{S}$ is constructed as follows. For each group $\mathcal{H}%
_{j}$, $j\in\mathbb{N}_{l}$, Simes test \citep{Simes:1986} to test the global
null ${H}_{j}^{\dagger}$ that $\mathcal{H}_{j}$ contains no false nulls is
applied at Type I error level $\xi=0.01$, and $\hat{S}$ contains each index $j^{\prime}$
for which ${H}^{\dagger} _{j^{\prime}}$ is rejected. We choose a weak signal
setting with $\mu_{\ast}=0.1$ and a relatively more stringent Type I error level
$\xi=0.01$ to make it less easy for the qGBH to estimate $S$ correctly, and
thus to illustrate the reduced power of GBH under a nontrivial sparse
configuration when $S$ has to be estimated.

\begin{figure}[t]
\centering
\includegraphics[height=0.4\textheight,width=0.85\textwidth]{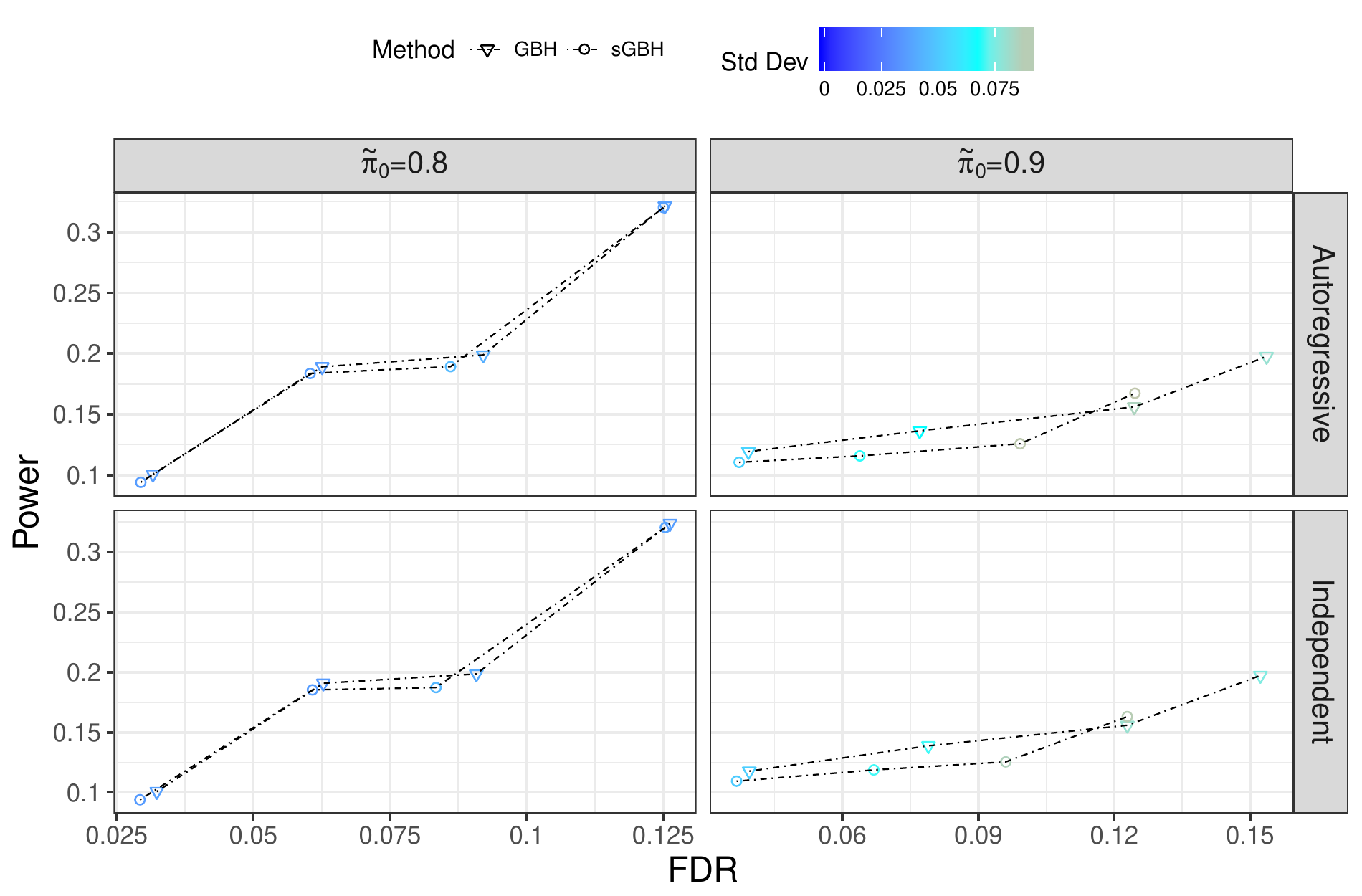}
\vspace{-0.5cm}\caption[power independence two-sided qsGBH]{FDRs and powers of
the quasi-adaptive sGBH (coded as \textquotedblleft sGBH\textquotedblright%
\ and as circle in the legend) and the adaptive GBH (coded as
\textquotedblleft GBH\textquotedblright\ and as downward triangle in the
legend) under independence and based on two-sided p-values when the subset $S$
of interesting groups is not correctly estimated under nontrivial sparse
configurations. Each type of points from left to right in each subfigure are
obtained successively under nominal FDR level $0.05,0.1,0.15$ and $0.2$. The
quasi-adaptive sGBH is conservative but less powerful than the oracle GBH.}%
\label{figPowerIndQ}%
\end{figure}

\autoref{figPowerIndQ} presents the powers and FDRs of the qGBH and the oracle
GBH for $4$ scenarios where $S$ has not always been correctly estimated. For
$m=4000$ and $\tilde{\pi}_{0} = 0.8$ or $0.9$, $S$ has only been correctly
estimated at most $91\%$ of the times, and the power loss of the oracle sGBH
can be as large as $5\%$. It can be perceived that the less frequently $S$ is
correctly estimated and the less accurately each $\pi_{j0}$ is estimated, the
more power loss the adaptive GBH will incur compared to its oracle version. We
will study data-adaptive sGBH in \autoref{secAdap} and empirically show
in \autoref{secSim} that it is more powerful than the adaptive GBH for
nontrivial sparse configurations.

\section{Adaptive versions of sGBH}

\label{secAdap}

Since in practice there is often prior information on the number of groups for
the hypotheses and their sizes, we assume that the partition $\mathcal{H}%
_{j}=\left\{  H_{j_{k}}:k\in G_{j}\right\}  $ for $j\in\mathbb{N}_{l}$ is known but that $S$ is unknown.
The adaptive sGBH is obtained by replacing $S$ by its estimate $\hat{S}$, setting each p-value in each group $G_{j}%
,j\notin\hat{S}$, to be infinite, replacing each $v_{j}$, $j\in\hat{S}$, by
its estimate $\hat{v}_{j}$, and applying BH to the set of weighted p-values $\left\{p_{j_{k}} \hat{v}_j: j \in \hat{S}, k \in G_j\right\}$.
The method used to estimate $S$ and the $v_{j}$'s determines the version of adaptive sGBH.

There are mainly two ways to obtain $\hat{v}_j, j \in S$. \cite{Hu:2010} obtained each $\hat{v}_{j}$
by respectively replacing $\pi_{j0}$ and $\pi_{0}$ in the definition of
$v_{j}$ by their estimates $\hat{\pi}_{j0}$ and $\hat{\pi}_{0}=m^{-1}%
\sum\nolimits_{j=1}^{l}\hat{\pi}_{j0}n_{j}$. We refer to these $\hat{v}_{j}$'s as
``plug-in'' weights and the resulting adaptive sGBH (or adaptive GBH) as the ``plug-in'' adaptive sGBH (or adaptive GBH).
When there are at least two groups, there does not seem to exist any
theoretical justification that the plug-in adaptive GBH is conservative non-asymptotically.
To overcome this issue, \cite{Nandi:2018} estimated each $v_{j}$ by
\begin{equation}
\hat{v}_{j}=\frac{\left(  n_{j}-R_{j}\left(  \lambda\right)  +1\right)  \left(
R\left(  \lambda\right)  +l-1\right)  }{m\left(  1-\lambda\right)
R_{j}\left(  \lambda\right)  },\label{eqe2a}%
\end{equation}
where $\lambda\in (0,1)$ is a tuning parameter, $R_{j}\left(  \lambda\right)  =\sum_{i\in G_{j}}\mathbf{1}_{\left\{
p_{i}\leq\lambda\right\}  }$ and $R\left(  \lambda\right)  =\sum_{i=1}%
^{m}\mathbf{1}_{\left\{  p_{i}\leq\lambda\right\}  }$. We refer to these weights as ``generic weights'' and the resulting
adaptive sGBH (or adaptive GBH) as the ``generic adaptive GBH'' (or generic adaptive GBH).
In fact, Theorem 2 of \cite{Nandi:2018} justifies the non-asymptotic
conservativeness of the generic adaptive GBH when p-values are independent and null p-values are uniformly distributed.

We point out that for the generic adaptive sGBH there is no need to estimate the $\tilde{\pi}_{0}$ in (\ref{eq10a1}),
and for the plug-in adaptive sGBH (and plug-in adaptive GBH) we will use the same estimator for each $\pi_{j0}$ unless otherwise noted.
Note that the plug-in adaptive sGBH makes no rejections when $\hat{\pi}_{0,\hat{S}}=1$, where
\[
\hat{\pi}_{0,\hat{S}}=\left(  \sum\nolimits_{j\in\hat{S}}\hat{\pi}_{j0}%
n_{j}\right)  \left(  \sum\nolimits_{j\in\hat{S}}n_{j}\right)  ^{-1}
\]
estimates $\tilde{\pi}_0$.

To investigate the conservativeness of the plug-in adaptive sGBH, we start
with proportion estimators and introduce the following definition.

\begin{definition}
\label{Def2}Let there be $m^{\prime}$ null hypotheses $\left\{  H_{i}^{\prime
}\right\}  _{i=1}^{m^{\prime}}$, for which $I_{0}^{\prime}$ is the index set
of true nulls and $p_{i}^{\prime}$ is the p-value associated with
$H_{i}^{\prime}$. Let $\pi_{0}^{\prime}$ be the proportion of true nulls for
$\left\{  H_{i}^{\prime}\right\}  _{i=1}^{m^{\prime}}$. An estimator $\hat
{\pi}_{0}^{\dag}$ of $\pi_{0}^{\prime}$ based on the p-values $\mathbf{p}%
^{\prime}=\left(  p_{1}^{\prime},\ldots,p_{m}^{\prime}\right)  $ is called
\textquotedblleft non-increasing in $\mathbf{p}^{\prime}$\textquotedblright%
\ if $\hat{\pi}_{0}^{\dag}$ is non-increasing in each $p_{i}^{\prime}$. Let
$\hat{\pi}_{0,k}^{\dag}$ be the estimator obtained by applying $\hat{\pi}%
_{0}^{\dag}$ to $\mathbf{p}_{0,k}^{\prime}=\left(  p_{1}^{\prime}%
,\ldots,p_{k-1}^{\prime},0,p_{k+1}^{\prime},\ldots,p_{m}^{\prime}\right)  $
for each $k\in I_{0}^{\prime}$. $\hat{\pi}_{0}^{\dag}$ is called
``reciprocally conservative'' (or has the
property of ``reciprocal conservativeness'') if
\begin{equation}
\mathbb{E}\left[  \left.  1\right/  \hat{\pi}_{0,k}^{\dag}\right]  \leq\left.
1\right/  \pi_{0}^{\prime}\text{ }\ \text{for each \ }k\in I_{0}^{\prime}.
\label{eqRConserve}%
\end{equation}

\end{definition}

Inequality (\ref{eqRConserve}) together with Jensen's inequality implies
$\mathbb{E}\left(  \hat{\pi}_{0,k}^{\dag}\right)  \geq\pi_{0}$. If $\hat{\pi
}_{0}^{\dag}$ is non-increasing, then $\hat{\pi}_{0,k}^{\dag}\leq\hat{\pi}%
_{0}^{\dag}$ almost surely for any $k\in I_{0}^{\prime}$, and for such
$\hat{\pi}_{0}^{\dag}$ reciprocal conservativeness implies conservativeness.
When p-values are independent but uniformly distributed under the true nulls,
reciprocal conservativeness has been observed and used by
\cite{Benjamini:2006}, \cite{Sarkar:2008b}, \cite{Blanchard:2009} and \cite{Chen:2016} to show
the conservativeness of adaptive FDR procedures, and examples of
non-increasing and reciprocally conservative estimators, including Storey's
estimator of \cite{Storey:2004}, are given by Corollary 13 of \cite{Blanchard:2009}. Specifically,
Storey's estimator is defined as%
\begin{equation}
\hat{\pi}_{0}^{\sharp}  =\frac{1}{m^{\prime}\left(  1-\lambda\right)
}+\frac{1}{m^{\prime}}\sum\limits_{i=1}^{m^{\prime}}\frac{\mathbf{1}_{\left\{  p^{\prime}_{i}%
>\lambda\right\}  }}{1-\lambda}\label{4A}%
\end{equation}
for a tuning parameter $\lambda\in (0,1)$.
However, these reciprocally conservative estimators are usually inconsistent. A consistent estimator will be discussed
in \autoref{SecEstpi0}.

We will provide simple conditions on the non-asymptotic conservativeness of
the plug-in adaptive sGBH. Let $m_{S}$ be the cardinality of $\tilde{\mathbf{p}}_{S}$
(and hence of $\mathbf{p}_{S}=\left\{  p_{j_{k}}:j\in S,k\in G_{j}\right\}  $),
and $\hat{\alpha}_{m}$ the FDR of the plug-in adaptive sGBH. We have

\begin{theorem}
\label{thmConserve}Consider a nontrivial sparse configuration where
$\tilde{\pi}_{0}\in\lbrack0,1)$ uniformly in $m$, and assume that $\left\{
p_{i}\right\}  _{i=1}^{m}$ are mutually independent. If each $\hat{\pi}_{j0}, j \in \hat{S}$ is
non-increasing and reciprocally conservative and $\Pr\left(  \hat{\pi}%
_{0,\hat{S}}<1\right)  >0$, then there exists a constant $\check{\pi}_{0}%
\in\lbrack0,1)$ such that $\Pr\left(  \hat{\pi}_{0,\hat{S}}\leq\check{\pi}%
_{0}\right)  >0$ and
\begin{equation}
\hat{\alpha}_{m}\leq\frac{\alpha}{m_{S}}\frac{1}{1-\check{\pi}_{0}}%
\sum\nolimits_{j\in S}n_{j}\left(  1-\pi_{j0}\right)  +\Pr\left(  \hat{S}\neq
S\right)  +\Pr\left(  \hat{\pi}_{0,\hat{S}}>\check{\pi}_{0}\right)  .
\label{bndwFDR}%
\end{equation}
If further $\hat{S}$ and $\hat{\pi}_{0,\hat{S}}$ consistently estimate $S$ and $\tilde{\pi}_{0}$ respectively, then
$\limsup_{m\rightarrow\infty}\hat{\alpha}_{m}\leq\alpha$. On the other hand,
if $\left\{p_{i}\right\}_{i=1}^{m}$ have the property of PRDS, $\hat{S}$ is consistent for $S$ and $\hat{\pi}_{j0}$ is consistent for $\pi_{j0}$ uniformly in
$j\in \hat{S}$ (without necessarily being non-increasing or reciprocally conservative), then $\limsup_{m\rightarrow\infty}\hat{\alpha}_{m}\leq\alpha$.
\end{theorem}

In \autoref{thmConserve}, the condition ``$\tilde{\pi}_{0}\in\lbrack0,1)$ uniformly in
$m$'' excludes the case ``$\tilde{\pi}_{0}=1$ for some $m$'', for which
the plug-in adaptive sGBH makes only false rejections (if any), and the assumption
$\Pr\left(  \hat{\pi}_{0,\hat{S}}<1\right)  >0$ excludes the case where the plug-in
adaptive sGBH makes no rejections and is conservative. \autoref{thmConserve}
does not require the number of groups $l$ to be constant in $m$, allows the
use of any non-increasing and reciprocally conservative estimator of the
proportion of true nulls for each interesting group, and accounts for effects of selecting
groups of hypotheses of interest. It generalizes Theorem 3 in \cite{Chen:2015discretefdr}. In particular, when all
groups are of interest, $\hat{S}=\left\{  1,\ldots,l\right\}  $ can be set,
and the upper bound $\eta_{m}$ on the right hand side of (\ref{bndwFDR})
reduces to that provided by Theorem 3 of \cite{Chen:2015discretefdr}.

The inequality (\ref{bndwFDR}) gives an integrated view on how grouping,
groupwise proportions, their estimates and selecting groups for weighting
affect the FDR of the plug-in adaptive sGBH. Specifically, $\eta_{m}$ bounds
$\hat{\alpha}_{m}$ from above, and $\eta_{m}\leq\alpha$ implies the
conservativeness of the procedure. So, for each $\alpha\in\left(  0,1\right)
$, the solution to $\eta_{m}\leq\alpha$ in terms of the accuracy of $\hat{S}$
and $\hat{\pi}_{0,\hat{S}}$, groupwise proportions $\left\{  \pi_{j0}\right\}
_{j\in S}$ and the constant $\check{\pi}_{0}$ corresponds to a setting where
the plug-in adaptive sGBH is conservative non-asymptotically (as in our simulation studies in \autoref{secSim}).

\subsection{Estimating the subset of interesting groups}

\label{SecEstS}

We deal with estimating $S$ under a nontrivial sparse configuration.
This is handled by the two-sided Kolmogorov-Smirnov (KS) test of
\cite{Kolmogorov:1933} and \cite{Smirnov:1939}. Specifically, for each
$j\in\mathbb{N}_{l}$, apply the KS test to all p-values in group $G_{j}$ to
test their uniformity, i.e., to test if they all follow the uniform
distribution, at some Type I error level $\beta >0$, and let $\hat{S}$
contain all $j$ such that the uniformity of p-values in group $G_{j}$ is rejected.
For each $j\in\mathbb{N}_{l}$, let $S_{j0}$ be the index set of true nulls among $\mathcal{H}_{j}$.
We have

\begin{lemma}
\label{ThmEstS}Consider a nontrivial sparse configuration. Assume $\left\{
p_{i}\right\}  _{i=1}^{m}$ are independent such that each $p_i$ is continuous and each $p_{i}, i\in
I_{0}$ is uniformly distributed. If $\lim
_{m\rightarrow\infty}\inf_{1\leq j\leq l}n_{j}=\infty$ and%
\begin{equation}
\lim_{m\rightarrow\infty}\inf_{j\in S}\frac{1}{\sqrt{n_{j}}\left(  1-\pi
_{j0}\right)  }\sup_{t\in\left[  0,1\right]  }\left\vert \sum\nolimits_{i\in
G_{j}\setminus S_{j0}}\left(  \mathbf{1}_{\left\{  p_{i}\leq t\right\}
}-t\right)  \right\vert =\infty\label{condEstS}%
\end{equation}
then $\hat{S}$ obtained by the KS test at any Type I error level $\beta>0$ satisfies
$\lim_{m\rightarrow\infty}\Pr(\hat{S}=S)=1$.
\end{lemma}

In \autoref{ThmEstS}, the condition (\ref{condEstS}) requires that the
empirical process associated with p-values corresponding to the false nulls in
each interesting group is well separately from the uniform distribution on
$\left[  0,1\right]  $. A condition that helps validate
(\ref{condEstS}) is%
\[
\limsup_{m\rightarrow\infty}\sup_{j\in S}\pi_{j0}=q\text{ for some }%
q\in\lbrack0,1)\text{,}%
\]
which requires that an interesting group always contain a positive proportion of false nulls.
Other tests, such as the higher criticism (HC) of \cite{Donoho:2004} and the
Simes test of \cite{Simes:1986}, on if an $\mathcal{H}_{j}$ contains all true
nulls can be used to estimate $S$. However, under a nontrivial sparse
configuration, we prefer to theoretically work under the conditions of \autoref{ThmEstS} and
use the KS test to ensure $\lim_{m\rightarrow\infty}\Pr(\hat{S}=S)=1$, so that
we may avoid dealing with the sparse case $1-\pi_{j0}\rightarrow0$ with $j\in
S$ better suited for HC and checking if the asymptotic power of Simes test
tends to $1$.

Even though the validity of the KS test was proved for a sequence of i.i.d.
observations, it essentially requires the supremum of a standardized empirical
process to converge to the supremum of a Brownian bridge. Thus, the KS test
may still perform well for strongly mixing random variables such as those
being autoregressive with order $1$. This has been observed for the Normal
means problem (to be defined in \autoref{SecEstpi0}) with an autoregressive
covariance matrix in the simulation study in \autoref{secSim}. However, the KS
test may be unreliable under moderately strong dependence, and due to its excellent power under independence it may be too stringent on
testing uniformity when applied in practice.

\subsection{Estimating the null proportions}

\label{SecEstpi0}

We discuss consistent proportion estimation, which is related to the
asymptotic conservativeness of the plug-in adaptive sGBH. There are only a few consistent proportion estimators
which are provided by \cite{swanepoel1999}, \cite{Meinshausen:2006},
\cite{Jin:2008} and \cite{Chen:2018a}. These estimators complement each other in terms of their
scopes of application. In particular, ``Jin's estimator'' of \cite{Jin:2008} has excellent performance for estimating proportions related to Normal means,
and its consistency has been extended by \cite{Chen:2018} to hold under a more general type
of dependence structure called ``principal covariance structure (PCS)''. In contrast, the
estimators of \cite{Meinshausen:2006} and \cite{swanepoel1999} require more stringent conditions than
Jin's estimator in order to be consistent.

We introduce the extended estimator of
\cite{Chen:2018} (referred to also as ``Jin's estimator'' for conciseness of notation). Let $\mathsf{Normal}\left(  \mathbf{a},\mathbf{A}\right)  $
denote the Normal distribution (or Normal random vector) with mean vector $\mathbf{a}$ and covariance
matrix $\mathbf{A}$. Let $\mathbf{y}=\left(  y_{1},\ldots,y_{n}\right)  $ with
$y_{i}\sim\mathsf{Normal}\left(  u_{i},s_{ii}\right)  $ have covariance
matrix $\mathbf{S}=\left(  s_{ij}\right)  $ and mean vector
$\boldsymbol{u}=\left(  u_{1},...,u_{n}\right)  $. Note that
$\mathbf{y}$ itself does not have to be a Normal random vector. Consider the
``Normal means problem'', i.e., simultaneously
testing the null $H_{i0}:u_{i}=0$ versus the alternative $H_{i1}:u_{i}%
\neq0$ for $1\leq i\leq n$. In this setting, $I_{0}^{\ast}=\left\{ i: 1 \le i \le n, u
_{i}=0\right\}  $, the proportion of zero Normal means $\pi_{0}^{\ast}$ is the ratio
of the number of zero $u_{i}$'s to $n$, and the proportion of nonzero Normal means is
$\pi^{\ast}=1-\pi_{0}^{\ast}$. The extended estimator $\hat{\pi}^{\ast}$ estimates $\pi^{\ast}$ and is
defined as follows. Let%
\[
\phi_{\mu,\sigma}\left(  x\right)  =\left(  \sqrt{2\pi}\sigma\right)
^{-1}\exp\left(  -2^{-1}\sigma^{-2}\left(  x-\mu\right)  ^{2}\right)
\]
for $\mu\in\mathbb{R}$ and $\sigma>0$, and $\omega$ be an even,
real-valued function defined on $\left(  -1,1\right)  $ that is non-negative and bounded by
some finite constant $K>0$ and Lebesgue integrates to
$1$. Define%
\[
\kappa_{\sigma}\left(  t;x\right)  =\int_{\left(  -1,1\right)  }\omega\left(
\zeta\right)  \exp\left(  2^{-1}t^{2}\zeta^{2}\sigma^{2}\right)  \cos\left(
t\zeta x\right)  d\zeta\text{,}%
\]
and
\begin{equation}
\varphi_{n}\left(  t;\mathbf{y}\right)  =\dfrac{1}{n}\sum_{i=1}^{n}\left(
1-\kappa_{\sqrt{s_{ii}}}\left(  t;y_{i}\right)  \right).
\label{eqd2}%
\end{equation}
Then $\varphi_{n}$ usually under-estimates $\pi^{\ast}$.

Let $\left\Vert \mathbf{S}\right\Vert _{1}=\sum_{i,j=1}^{n}\left\vert
\mathbf{S}\left(  i,j\right)  \right\vert $ and the big $O$ notation be
``Landau's big O''. When each pair of distinct
entries of $\mathbf{y}$ is bivariate Normal and $\left\{  y_{i}\right\}
_{i=1}^{n}$ have a PCS such that%
\begin{equation}
n^{-2}\left\Vert \mathbf{S}\right\Vert _{1}=O(n^{-\delta
})\ \ \text{for some }\delta>0,\label{eqd1}%
\end{equation}
we can set%
\begin{equation}
\hat{\pi}^{\ast}=\varphi_{n}\left(  \sqrt{2\gamma\log n};\mathbf{y}\right)  \text{
for some }\gamma\in(0,2^{-1}\delta].\label{eqd3}%
\end{equation}
Then, $\hat{\pi}_{0}^{\ast}=1-\hat{\pi}^{\ast}$ estimates $\pi_{0}^{\ast}$. From Theorem 1 and the
proof of Theorem 2 of \cite{Chen:2018}, we can see that, $\hat{\pi}_{0}^{\ast}$
consistently estimates $\pi_{0}^{\ast}$ under PCS when $\pi_{0}^{\ast}\in(0,1]$, $\sup_{n\geq1}\max_{1\leq i\leq n}s_{ii}\leq1$ and
\begin{equation}
\lim_{n\rightarrow\infty}\sqrt{2\gamma\ln n}\min\left\{  \left\vert u
_{j}\right\vert :u_{j}\neq0, 1 \le j \le n \right\}  =\infty.\label{eqd4}%
\end{equation}
We refer interested readers to \cite{Jin:2008} and \cite{Chen:2018} for the
excellent empirical performances of $\hat{\pi}^{\ast}$ and $\hat{\pi}_{0}^{\ast}$ under PCS and various
sparse settings. Note that, when $\mathbf{y}\sim\mathsf{Normal}\left(
\boldsymbol{u},\mathbf{S}\right)  $ has a PCS with $\max_{1\leq i\leq
n}s_{ii}\leq1$, $\mathbf{S}$ being a correlation
matrix represents the most difficult case of estimating $\pi^{\ast}$ among different types of $\mathbf{S}$; see the
discussion right after Theorem 2 of \cite{Chen:2018}. The simulation study in
\autoref{secSim} for the Normal means problem sets $\mathbf{S}$ to be a
correlation matrix.

We remark on the relationship between PRDS and PCS. Consider a variant of the
Normal means problem where $H_{i0}:u_{i}=0$ and $H_{i1}:u_{i}>0$ for each
$1\leq i\leq n$. If $\mathbf{y}\sim\mathsf{Normal}\left(  \boldsymbol{u
},\mathbf{S}\right)  $ has a PCS and $\mathbf{S}$ has
nonnegative entries, then the distribution of $\mathbf{y}$ is PRDS on $I_{0}^{\ast}$,
and so are the one-sided p-values $1-\Phi\left(  y_{i}\right)  $
for $1\leq i\leq n$, where $\Phi$ is the CDF for $\mathsf{Normal}\left(  0,1\right)  $; see \cite{Benjamini:2001} for
a justification on this. In other words, under these settings PRDS and PCS are
compatible with each other to enable the theory presented by
\autoref{thmConserve} on asymptotic FDR control of the plug-in adaptive sGBH.

\section{Simulation study}

\label{secSim}

We will employ Storey's estimator or Jin's estimator to compare the performances of the plug-in adaptive sGBH and plug-in adaptive sGBH, and also compare the performances of
the generic adaptive GBH and generic adaptive sGBH. Let $\mathbf{z}\sim\mathsf{Normal}\left(  \boldsymbol{\mu},\mathbf{\Sigma}\right)  $, where $\mathbf{z}=\left(  z_{1},\ldots,z_{m}\right)  $, $\mathbf{\Sigma}=\left( \sigma_{ij}\right)  $ and
$\boldsymbol{\mu}=\left(  \mu_{1},...,\mu_{n}\right)  $.  We consider the Normal means problem under
PCS where $\mathbf{\Sigma}$ is a correlation matrix and
for each $H_{i0}:\mu_{i}=0$, the two-sided p-value $p_{i}%
=2\Phi\left(  -\left\vert z_{i}\right\vert \right)  $ and the one-sided p-value
${p}_{i}=1-\Phi\left(  z_{i}\right)  $. Since sGBH coincides
with GBH when $S=\varnothing$ or $S=\mathbb{N}_{l}$, a sparse configuration with
$S\neq\varnothing$ and $S \ne \mathbb{N}_{l}$ will be used for the hypotheses.

\subsection{Simulation design}

\label{simDesign}

For $\mathbf{z}\sim\mathsf{Normal}\left(  \boldsymbol{\mu},\boldsymbol{\Sigma
}\right)  $ with $\boldsymbol{\Sigma}$ being a correlation matrix, we consider
$6$ values for $m$ as $4\times10^{3}$, $10^{4}$, $2\times10^{4}$,
$4\times10^{4}$, $8\times10^{4}$ and $10^{5}$, and $2$ types of dependence
encoded by $\boldsymbol{\Sigma}$ that satisfy PCS defined by (\ref{eqd1}).
Specifically, $\boldsymbol{\Sigma}=\mathsf{diag}\left(  \boldsymbol{\Sigma
}_{1},\boldsymbol{\Sigma}_{2},\boldsymbol{\Sigma}_{3},\boldsymbol{\Sigma}%
_{4}\right)  $ is block diagonal with $4$ blocks of equal sizes and
$\boldsymbol{\Sigma}_{k}=\left(  \sigma_{ij,k}\right)  $ for $1\leq k\leq4$,
set as follows:

\begin{itemize}
\item \textquotedblleft Independent\textquotedblright: $\sigma_{ij}=0$ when
$i\neq j$, i.e., $\boldsymbol{\Sigma}$ is the identity matrix.

\item \textquotedblleft Autoregressive\textquotedblright: $\sigma_{ij,k}%
=\rho_{k}^{\left\vert i-j\right\vert }1_{\left\{  i\neq j\right\}  }$ with
$\rho_{k}=0.1k$ for $k=1,\ldots,4$. Each $\boldsymbol{\Sigma}_{k}$ is the
autocorrelation matrix of an autoregressive model of order $1$, such that%
\begin{equation}
4^{-2}m^{-2}\left\Vert \boldsymbol{\Sigma}_{k}\right\Vert _{1}=\frac{1}%
{16}\frac{1+\rho_{k}}{1-\rho_{k}}m^{-1}+O\left(  m^{-2}\right)  . \label{eq2}%
\end{equation}

\end{itemize}

\noindent The Autoregressive dependence given above is strongly mixing and
interpolates block dependence and short-range dependence. Note that
$\boldsymbol{\Sigma}$ itself encodes block dependence. Since each type of
$\boldsymbol{\Sigma}$ has nonnegative entries, $\mathbf{z}$ satisfies PRDS on
$I_{0}$. By the discussion at the end of \autoref{SecEstpi0}, we see that the
one-sided p-values satisfy PRDS on $I_{0}$ whereas the two-sided ones may not.

For sGBH, the configuration for $\left\{  H_{i}\right\}  _{i=1}^{m}$ is as
follows. There are $4$ groups of hypotheses%
\[
\mathcal{H}_{k}=\left\{  H_{i}:i=4^{-1}m\left(  k-1\right)  +1,\ldots
,4^{-1}mk\right\}
\]
for $k=1,\ldots,4$, such that $S=\left\{  1\right\}  $ and $\tilde{\pi}%
_{0}=\pi_{10}=0.7,0.8$ or $0.9$. Namely, each $\mathcal{H}_{k}$ matches the
corresponding $\boldsymbol{\Sigma}_{k}$ and only $\mathcal{H}_{1}$ contains
false nulls. This particular partition for the hypotheses is not tailored for
the adaptive sGBH to be more powerful than the adaptive GBH but is meant for
easy computer simulation. However, it will reveal a general phenomenon about
GBH in non-asymptotic settings as we will explain in \autoref{simRes}.

The nonzero $\mu_{i}$'s are generated independently such that their absolute
values $|\mu_{i}|$ are from the uniform distribution on the compact interval
$\left[  0.6,3.6\right]  $ but each $\mu_{i}$ has probability $0.5$ to be
negative or positive. Let $\mu_{\ast}=\min\left\{  \left\vert \mu
_{i}\right\vert :\mu_{i}\neq0\right\}  $, and recall the sufficient condition
(\ref{eqd4}) needed to ensure the consistency of Jin's estimator $\hat{\pi}_0^{\ast}$ when it is applied to estimate the $\pi_{j0}$'s. Here $\mu_{\ast
}=0.6$, and (\ref{eqd4}) is satisfied. Further, $\hat{\pi}_0^{\ast}$ with $\gamma=0.5$
is used to estimate $\pi_{j0}$ for each $j\in\mathbb{N}_{l}$. Note that choosing $\gamma=2^{-1}\delta$, where $\delta$
is the ``PCS index'' appearing in (\ref{eqd1}) and $\delta=1$ here for each $\boldsymbol{\Sigma}_k%
$, leads to relatively fast convergence of $\hat{\pi}_{0}^{\ast}$ to achieve
consistency but may cause $\hat{\pi}_{0}^{\ast}$ not to have relatively small
variance.

There are $4$ nominal FDR\ levels, i.e., $\alpha=0.05,0.1,0.15$ or $0.2$. The
simulation is implemented by independently repeating $200$ times each
experiment determined by the quintuple $\left(  \alpha,m,\tilde{\pi}%
_{0},\boldsymbol{\Sigma},p\right)  $ for a total of $288=144\times2$
scenarios, where $p$ denotes a one-sided or two-sided p-value. Storey's estimator is implemented by the \textsf{pi0est}\textrm{
}function with parameter `\textsf{smoother}\textrm{'} from the
\textsf{q-value} package. For the generic adaptive sGBH and generic adaptive GBH, the weights
in (\ref{eqe2a}) are obtained with $\lambda=0.25$, $0.5$ or $0.75$. When the KS test is used to estimate $S$, it is implemented at
Type I error $\beta=0.025$.

\subsection{Simulation results}

\label{simRes}

We will visualize major results based on two-sided p-values in the main text
but gather in the supplementary material those based on one-sided p-values.
To measure the power of an FDR procedure, we use the expectation of the true
discovery proportion (TDP), defined as the ratio of the number of rejected
false nulls to the total number of false nulls. Note that FDR is the
expectation of the false discovery proportion (FDP), defined as the ratio of
the number of rejected true nulls to the total number of rejections. We also
report but do not focus on the standard deviations of the FDP and TDP since
smaller standard deviations for these quantities mean that the corresponding
procedure is more stable in FDR and power.

\begin{figure}[H]
\centering
\includegraphics[height=0.87\textheight,width=\textwidth]{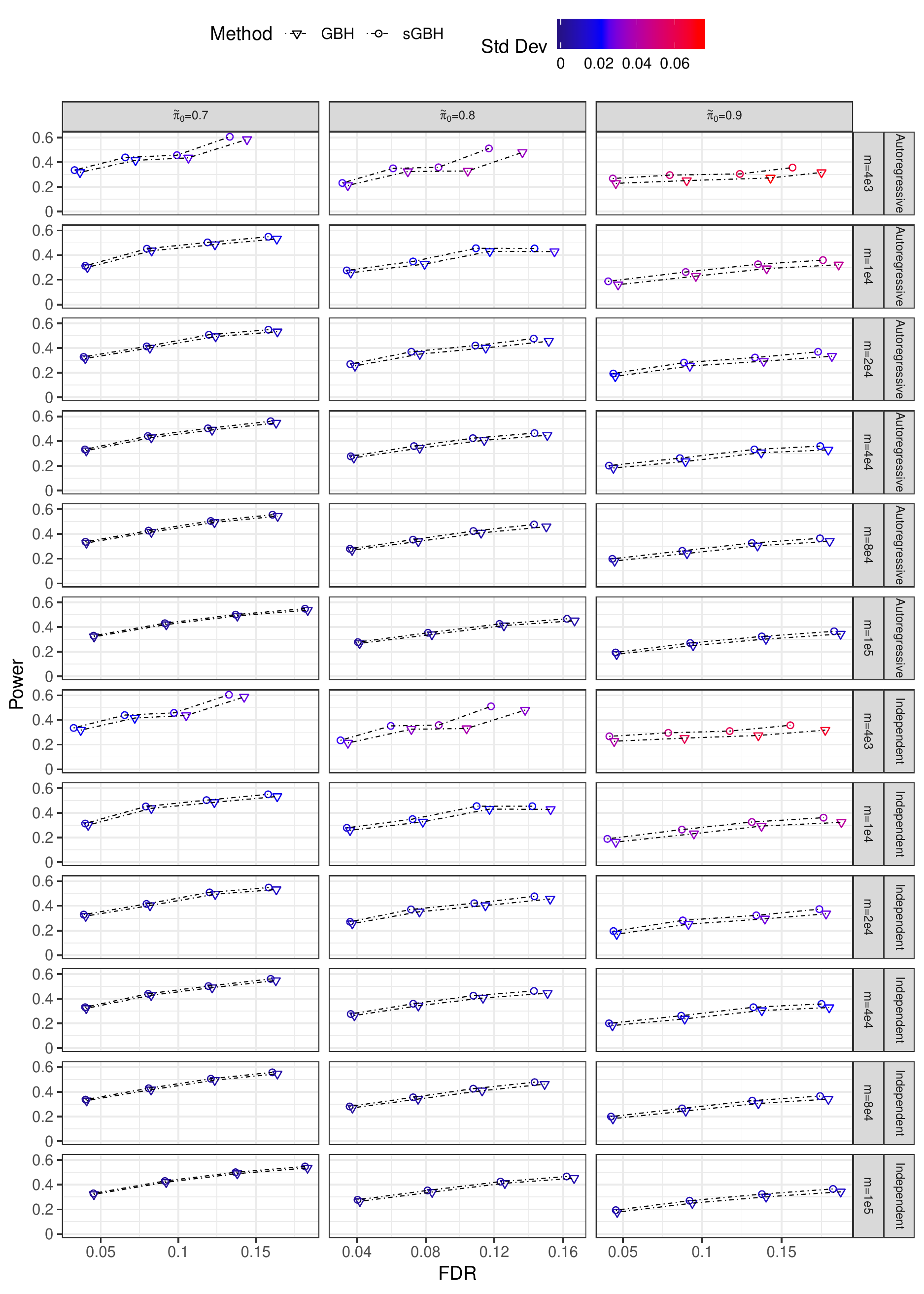}
\vspace{-0.7cm}\caption[power independence two-sided jin est]{FDRs and powers
of the plug-in adaptive sGBH (``sGBH'') and GBH (``GBH'') based on two-sided p-values. Each type of points from left to
right in each subfigure are obtained successively under nominal FDR level
$0.05,0.1,0.15$ and $0.2$. The color legend ``Std Dev'' is the standard deviation of the FDP. The KS test has been used to identify interesting groups, and Jin's estimator to estimate the null
proportions.}%
\label{figPowerJin2Side}%
\end{figure}

\begin{figure}[H]
\centering
\includegraphics[height=0.87\textheight,width=0.95\textwidth]{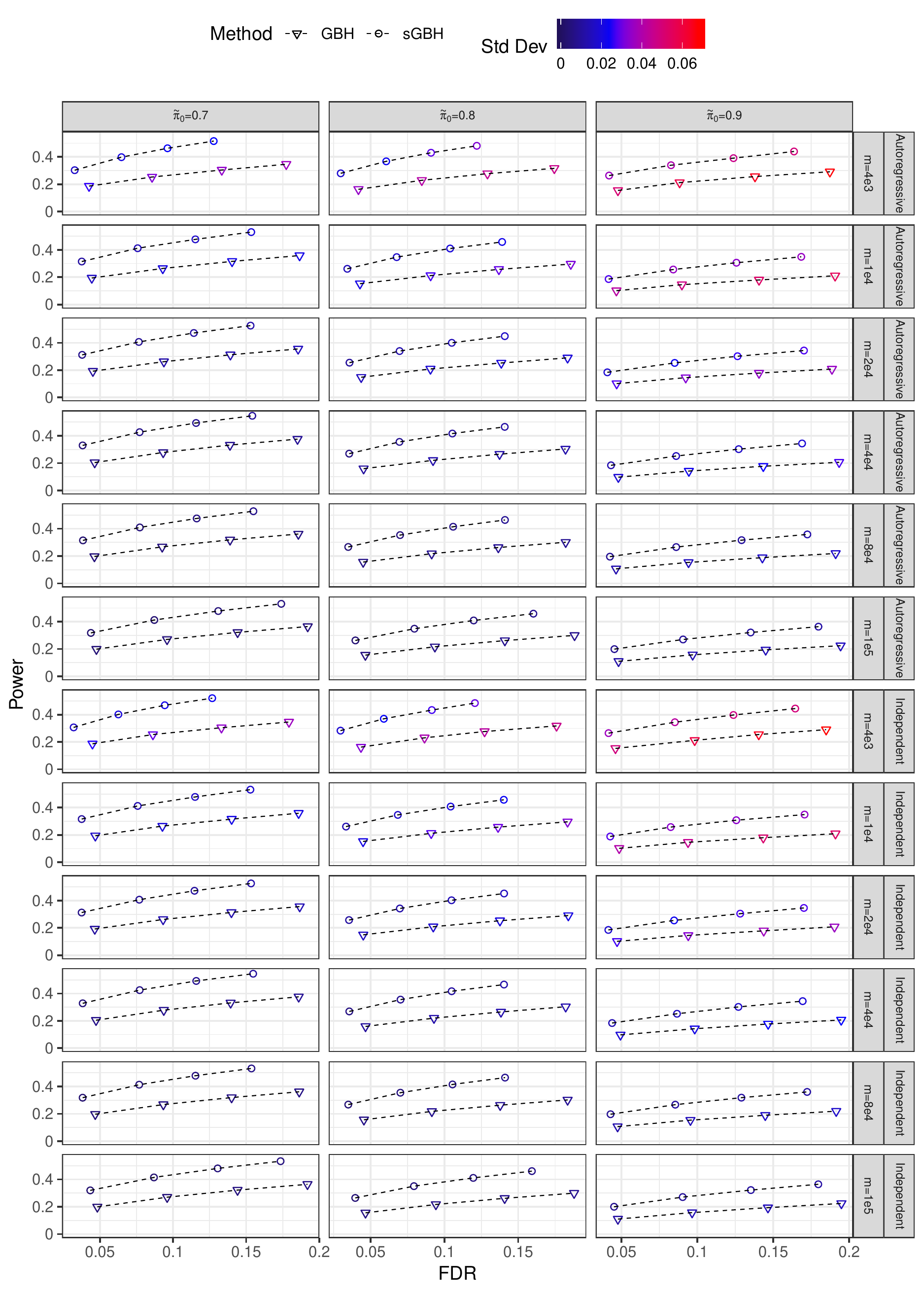}
\vspace{-0.5cm}\caption[power new, 2-side]{FDRs and powers of the generic adaptive
sGBH (``sGBH'') and GBH (``GBH'') based on two-sided p-values and with tuning parameter $\lambda=0.5$. Each type of points from left to right in each subfigure
are obtained successively under nominal FDR level $0.05,0.1,0.15$ and $0.2$. The color legend ``Std Dev'' is the standard deviation of the FDP. The KS test has been used to identify interesting groups.}%
\label{figPowerNew2Side}%
\end{figure}

\begin{figure}[H]
\centering
\includegraphics[height=0.87\textheight,width=.95\textwidth]{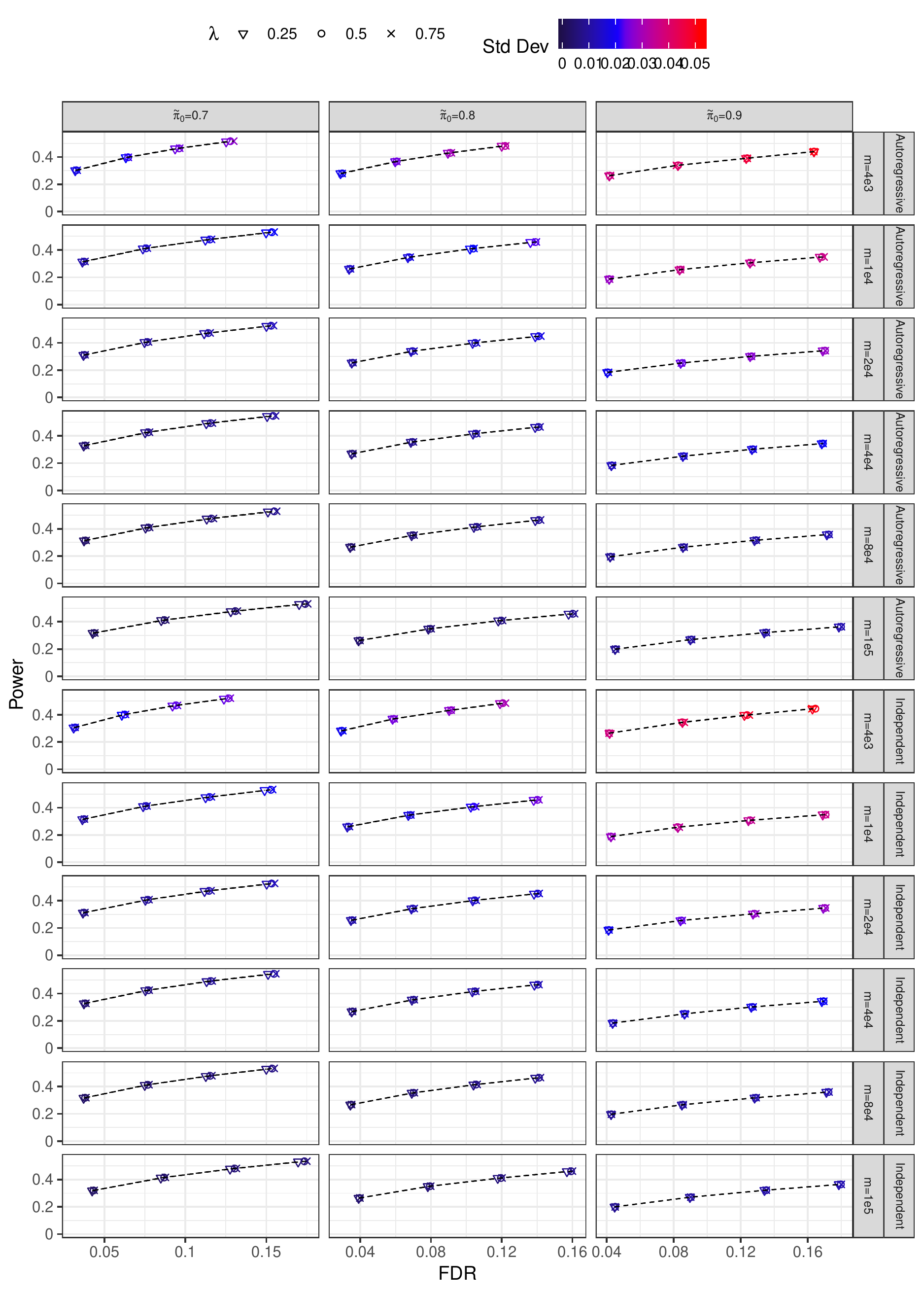}
\vspace{-0.5cm}\caption[power new, 2-sided, lambda]{FDR and power of the generic
adaptive sGBH (``sGBH") based on two-sided p-values and as the tuning
parameter $\lambda$ (shown in the legend) ranges in $\{0.25,0.5,0.75\}$. Each
type of points from left to right in each subfigure are obtained successively
under nominal FDR level $0.05,0.1,0.15$ and $0.2$. The color legend ``Std Dev'' is the standard deviation of the FDP. The KS test has been used to identify interesting groups.}%
\label{figPowerNewLambda2side}%
\end{figure}

\autoref{figPowerJin2Side} presents the FDRs and powers
of the plug-in adaptive sGBH and adaptive GBH that employ Jin's estimator, \autoref{figPowerNew2Side} those of the generic adaptive sGBH and adaptive GBH with tuning parameter $\lambda=0.5$,
and \autoref{figPowerNewLambda2side} those of the generic adaptive sGBH when the tuning parameter $\lambda$ (appearing in (\ref{eqe2a})) ranges in $\{0.25,0.5,0.75\}$, all based on two-sided p-values. For notational simplicity, the ``adaptive GBH (or adaptive sGBH)'' refers to the two versions, the plug-in and generic, adaptive GBH (or adaptive sGBH).
The following can be observed: (i) the adaptive sGBH is conservative and always has smaller FDR than the adaptive GBH
for all scenarios; (ii) the adaptive sGBH is more powerful than the adaptive GBH for all scenarios
related to two-side p-values, and it is so for one-sided p-values except when
$\tilde{\pi}_{0} = 0.9$ and $m=4000$; (iii) as the number of hypotheses
increases, the improvement in power of the plug-in adaptive sGBH upon the plug-in adaptive GBH
tends to decrease; (iv) the FDR and power of the generic adaptive sGBH change little even if we change $\lambda$; (v) for two-sided p-values, the plug-in adaptive sGBH with Jin's estimator has similar power to that of
the plug-in adaptive sGBH with Storey's estimator but is more powerful than the generic adaptive sGBH. Similar power characteristics can be
observed for the plug-in adaptive GBH for two-sided p-values;
(vi) the plug-in adaptive GBH with Storey's estimator can be anti-conservative under independence, as indicated by
the scenario $\tilde{\pi}_0 = 0.9$ and $m=4000$, and it can have very low power for one-sided p-values (since for a point null a one-sided p-value
corresponds to a misspecified test and each of the means, i.e., $\mu_i$'s, has equal probability to be positive or negative in our simulation). The explanations for these observations are provided below.

Recall $m_{S}$ as the cardinality of $\mathbf{p}_{S}=\left\{  p_{j_{k}}:j\in
S,k\in G_{j}\right\}  $. Under a nontrivial sparse configuration, we have $1>\pi
_{0}\geq\tilde{\pi}_{0}$, $m\geq m_{S}$ and
\begin{equation}
\frac{\left(  1-\tilde{\pi}_{0}\right)  m_{S}}{\left(  1-\pi_{0}\right)  m}=1.
\label{eqg2}%
\end{equation}
First, consider the two oracle procedures. Then
\begin{equation}
{p_{j_{k}}^{\ast}=}\frac{{p}_{j_{k}}\pi_{j0}}{1-\pi_{j0}}\left(  1-\pi
_{0}\right)  \text{ \ and \ }{\tilde{p}_{j_{k}}=}\frac{{p}_{j_{k}}\pi_{j0}%
}{1-\pi_{j0}}\left(  1-\tilde{\pi}_{0}\right)  \label{eqg1}%
\end{equation}
for $j\in S$ and $k\in G_{j}$ respectively for the oracle GBH and the oracle
sGBH. So%
\begin{equation}
{p_{j_{k}}^{\ast}\geq\tilde{p}_{j_{k}}}\text{ \ for each }j\in S\text{ and
}k\in G_{j}. \label{eqg3}%
\end{equation}
Since neither of the oracle procedures rejects any hypothesis $H_{j}$ with
$j\notin S$ and $k\in G_{j}$, the identities (\ref{eqg2}) and (\ref{eqg3})
force them to reject the same set of hypotheses. This is what
\autoref{ThmOracle} asserts.

Now consider the adaptive versions of the two oracle procedures when the
adaptive sGBH correctly estimates $S$. Set
\[
\hat{\pi}_{0,S}=\left(  \sum\nolimits_{j\in S}\hat{\pi}_{j0}n_{j}\right)
\left(  \sum\nolimits_{j\in S}n_{j}\right)  ^{-1},%
\]
which estimates $\tilde{\pi}_0$. Then, for $j\in S$ and $k\in G_{j}$,
${p_{j_{k}}^{\ast}}$ and ${\tilde{p}_{j_{k}}}$ respectively become
\begin{equation}
{\hat{p}_{j_{k}}^{\ast}=}\frac{{p}_{j_{k}}\hat{\pi}_{j0}}{1-\hat{\pi}_{j0}%
}\left(  1-\hat{\pi}_{0}\right)  \text{ \ and \ }{\hat{\tilde{p}}_{j_{k}}%
=}\frac{{p}_{j_{k}}\hat{\pi}_{j0}}{1-\hat{\pi}_{j0}}\left(  1-\hat{\pi}%
_{0,S}\right).  \label{eqg4}%
\end{equation}
Note that $\pi_{j0}=1$ for $j\notin S$. If $\hat{\pi
}_{j^{\prime}0}\neq1$ when $\pi_{j^{\prime}0}=1$ for some $j^{\prime}\notin
S$, then the plug-in adaptive GBH will likely reject a hypothesis, say, $H_{j_{k}^{\prime}}$,
from the group $G_{j^{\prime}}$, potentially leading to increased FDR and
anti-conservativeness. However, when the adaptive sGBH correctly estimates
$S$, it will never reject this $H_{j_{k}^{\prime}}$ and will never reject any
$H_{j}$ with $j\notin S$ and $k\in G_{j}$, likely leading to conservativeness and
potentially smaller FDR than the adaptive GBH. Further, due to the potential
inconsistency of proportion estimation in non-asymptotic settings, the order
between ${p_{j_{k}}^{\ast}}$ and ${\tilde{p}_{j_{k}}}$ for $j\in S$ and $k\in
G_{j}$ given by (\ref{eqg3}) for the oracles no longer necessarily holds for
${\hat{p}_{j_{k}}^{\ast}}$ and ${\hat{\tilde{p}}_{j_{k}}}$, and (\ref{eqg2})
becomes $\left(  1-\hat{\pi}_{0,S}\right)  m_{S}\left(  \left(  1-\hat{\pi
}_{0}\right)  m\right)  ^{-1}$, not necessarily being $1$. This allows the plug-in
adaptive sGBH to be uniformly more powerful than the plug-in adaptive GBH in
non-asymptotic settings. However, as $m$ increases and becomes sufficiently
large, the effect of asymptotic theory comes into play, the estimates of $S$
and each $\pi_{j0}$ become more accurate, and eventually both plug-in adaptive
procedures converge to their oracle versions, having identical performance.

We have observed that the subset $S$ of interesting groups has been correctly
estimated in each repetition of each simulation scenario except when
$\tilde{\pi}_{0} = 0.9$ and $m=4000$ and one-sided p-values were used, and for a few repetitions of some experiments
one of the uninteresting groups has been identified as an interesting group. In other words, it is easier
to consistently estimate $S$ than to consistently estimate $\pi_{j0}$ for each
$j\in\mathbb{N}_{l}$. So, under a nontrivial sparse configuration, the
adaptive sGBH tends to perform better than the adaptive GBH.
The estimated groupwise proportions for the
uninteresting groups and the estimate of $\tilde{\pi}_{0}$ are provided in the supplementary material.
Regardless of if Jin's estimator or Storey's estimate is used, for each
uninteresting group the frequency of its estimated null proportion being $1$ is
considerably less than $1$, and the estimated $\pi_0$ and $\tilde{\pi}_0$ are often
smaller than $1$. Therefore, the plug-in weights for these uninteresting groups are often finite, and the plug-in adaptive GBH tends to make more false discoveries and
be less powerful than the plug-in adaptive sGBH.

Further, we explain why for one-sided p-values, the plug-in adaptive sGBH is less
powerful than the plug-in adaptive GBH when $\tilde{\pi}_{0}=0.9$ and $m=4000$. When
$\mu_{i}\neq0$, its associated one-sided p-value $\check{p}_{i}=1-\Phi\left(
z_{i}\right)  $ tends to be larger when $\mu_{i}<0$ than it is when $\mu_{i}
\ge0$. So, many $\check{p}_{i}$'s will be relatively large when their
corresponding $\mu_{i}$'s are negative and small, there is a power loss when
conducting multiple testing based on $\check{p}_{i}$'s, the KS test will not
have enough power based on $\check{p}_{i}$'s when $m$ is small, and the
proportion estimators will have inflated biases. In fact, in this scenario, the subset $S$ of
interesting groups is only correctly estimated for at most $39\%$ of the times,
$\hat{\pi}_{0,\hat{S}}$ over-estimates $\tilde{\pi}_0$ and is often close to $1$ when $S$ is correctly estimated,
$\hat{\pi}_{0}$ is often larger than $\hat{\pi}_{0,\hat{S}}$, and neither $\hat{\pi
}_{0,\hat{S}}$ nor $\hat{\pi}_{0}$ is identically $1$.
This leads the adaptive plug-in sGBH to make less rejections and hence be less powerful than the plug-in adaptive GBH
since the latter is applied to all p-values whereas the former only to those in the estimated interesting groups.

Finally, we explain (a) why the two versions of the plug-in adaptive sGBH (or GBH) based on Jin's estimator and Storey's estimator have similar powers and (b) why the plug-in adaptive sGBH (or GBH) is more powerful than the generic adaptive sGBH (or GBH). Even though Jin's estimator provides more accurately estimates of $\tilde{\pi}_0$ than Storey's estimator, it estimates the groupwise proportions less accurately than Storey's estimator. This explains (a). On the other hand,
the plug-in weights are more aggressive estimates than the generic weights in terms of estimating the oracle weights $\{v_j\}_{j=1}^l$. So, the plug-in adaptive sGBH (or GBH) is often more powerful than the generic adaptive sGBH (or GBH), even though at the risk of being anti-conservative non-asymptotically. This explains (b).

\begin{figure}[H]
\centering
\includegraphics[height=0.87\textheight,width=.95\textwidth]{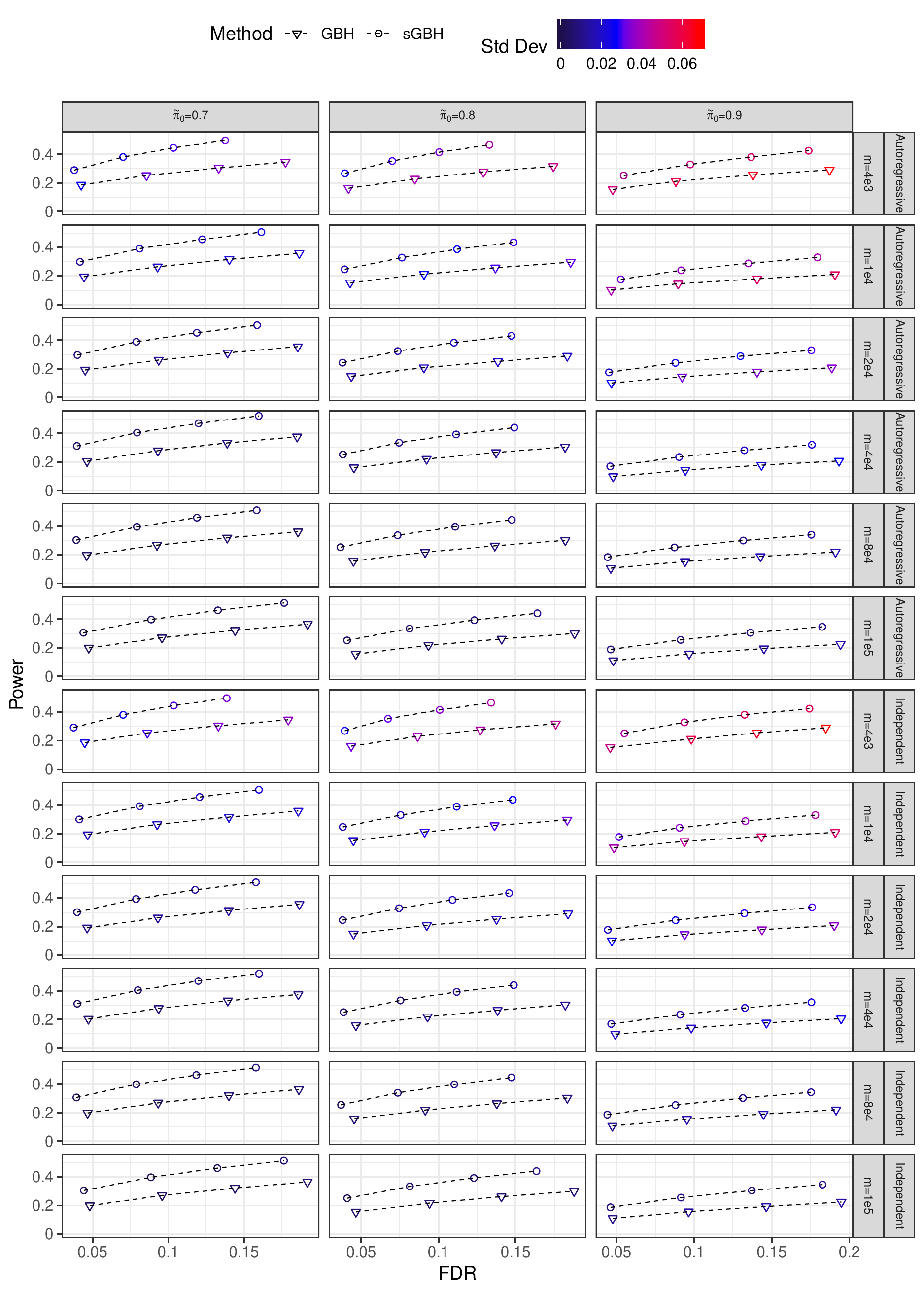}
\vspace{-0.5cm}\caption[power new, 2-sided, simes]{FDRs and powers
of the generic adaptive sGBH (``sGBH'') and GBH (``GBH'') based on two-sided p-values and tuning parameter $\lambda=0.5$. Each type of points from left to
right in each subfigure are obtained successively under nominal FDR level
$0.05,0.1,0.15$ and $0.2$. The color legend ``Std Dev'' is the standard deviation of the FDP. Simes test has been used to identify interesting groups at Type I error level $\xi=0.1$.}%
\label{figPowerNewSimes2side}%
\end{figure}

\begin{figure}[H]
\centering
\includegraphics[height=0.87\textheight,width=.95\textwidth]{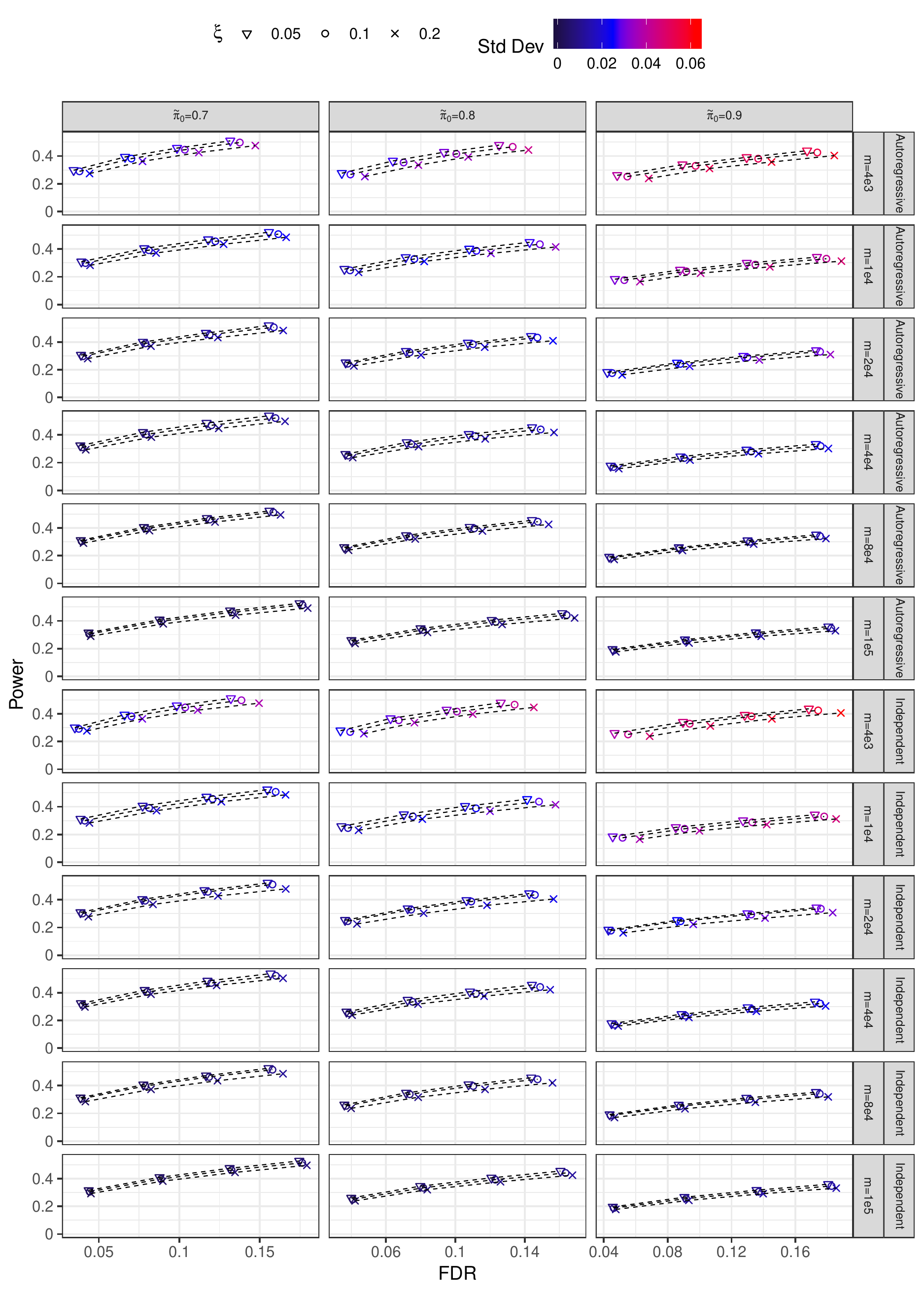}
\vspace{-0.5cm}\caption[power new, 2-sided, simes, xi]{FDRs and powers
of the generic adaptive sGBH (``sGBH'') and GBH (``GBH'') based on two-sided p-values, with tuning parameter $\lambda=0.5$ and as the Type I error
level $\xi$ (shown in the legend) of Simes test ranges in $\{0.05,0.1,0.2\}$. Each type of points from left to
right in each subfigure are obtained successively under nominal FDR level
$0.05,0.1,0.15$ and $0.2$. The color legend ``Std Dev'' is the standard deviation of the FDP. Simes test has been used to identify interesting groups.}%
\label{figPowerNewSimesXi2side}%
\end{figure}

We also examined the generic adaptive sGBH under the same settings given in \autoref{simDesign} but used Simes test to identify the set $S$ of interesting groups at Type I error level $\xi = 0.05, 0.1$ or $0.2$. For each repetition of each of the $288$ experiments and each of the three $\xi$ values, $S$ was correctly identified. However, similar to the KS test, for a few repetitions of some experiments one of the uninteresting groups has been identified as an interesting group. \autoref{figPowerNewSimes2side} presents the comparison between the generic adaptive sGBH and GBH, and \autoref{figPowerNewSimesXi2side} the performance of the generic adaptive sGBH as $\xi$ changes, both for two-sided p-values. The generic adaptive sGBH is conservative and always more powerful than the generic adaptive GBH for all $3$ values of $\xi$, and the two procedures have competitive FDRs. The former procedure has more power improvement over and has smaller FDR than the latter when $\xi$ is smaller. This is reasonable since the smaller $\xi$ is, the less likely an uninteresting group that has no false nulls will be identified as an interesting group. We did not examine the generic adaptive sGBH that employs Simes test to select groups of hypotheses with a fixed Type I error level $\xi$ but as the tuning parameter $\lambda$ changes since we suspect that the FDR and power performances of the procedure under this setting should be similar as $\lambda$ changes, in view of the performances of the generic adaptive sGBH that employs the KS test at a fixed Type I error level but as $\lambda$ changes.

\section{An application of sGBH}

\label{secApp}

We apply the adaptive sGBH to the prostate cancer data set of
\cite{Singh:2002}, with comparison to the adaptive GBH. The data set contains
expressions of approximately $12600$ genes from $52$ patients with prostate
tumors and $50$ normal specimens, and the target is to identify genes that are
differentially expressed between the two biological conditions. Under each
biological condition, each gene expression is modelled by a Normal
random variable. The detailed analysis is given below.

For gene $i$, p-value $p_{i}$ from a two-sided two-sample t-test and the
z-score $z_{i}=\Phi^{-1}(p_{i})$ are obtained. A total of $4374$ hypotheses
are selected and partitioned into $3$ groups as follows: $\mathcal{H}_{1}$
contains hypotheses whose associated p-values are bigger than $0.7$ and has
$1374$ elements; $\mathcal{H}_{2}$ contains $1500$ hypotheses that are
randomly sampled from hypotheses whose associated p-values are between $0.15$
and $0.7$; $\mathcal{H}_{3}$ contains $1500$ hypotheses that are randomly
sampled from hypotheses whose associated p-values are less than $0.15$. For
such a configuration, $\mathcal{H}_{1}$ likely will contain many more true
nulls than false nulls, and $\mathcal{H}_{3}$ many false nulls than true
nulls. The PCS index $\delta$ defined by (\ref{eqd1}) for the z-scores are
respectively $0.1826$, $0.0880$ and $0.1546$, revealing dependencies much
stronger than covered by our simulation study. In fact, the KS test at Type I error level $0.1$
asserts that each group contains some false nulls. So, instead of the KS test, we
use Simes test to select groups of interesting hypotheses as done in
\autoref{secQGBH} but at Type I error level $0.1$, which claims $\mathcal{H}_{2}$ as
interesting and $\hat{S}=\left\{  2\right\}  $. The estimated groupwise
proportions are $\hat{\pi}_{10}=0.9317$, $\hat{\pi}_{20}=0.8171$ and $\hat
{\pi}_{30}=1$, where $\gamma= (\delta- 0.0001)/2$ is used for Jin's
estimator. This gives $\hat{\pi}_{0}=0.9158$ and $\hat{\pi}_{0,\hat{S}}%
=\hat{\pi}_{20}=0.8171$. At nominal FDR level $0.05$, the plug-in adaptive sGBH claims
that $759$ are differentially expressed as compared to $487$ claimed by the plug-in
adaptive GBH, showing a considerable improvement.

We are well aware that different configurations may affect the performances of
the two plug-in adaptive procedures. So, we have tried other schemes of selecting from the
$12600$ hypotheses and then partitioning the selected hypotheses with
configurations different than the one given above. However, for all schemes we
have tried the plug-in adaptive sGBH had no less rejections than the plug-in adaptive GBH, and
for some the former had more rejections than the latter, all at the same FDR
level. This, together with \autoref{PropQgbh}, the discussion right after it,
and the simulation results in \autoref{secSim}, suggests that with the
same partition of hypotheses and the same proportion estimators, the plug-in adaptive
sGBH usually does not have less rejections than the plug-in adaptive GBH. Maintaining $\hat{S}=\left\{  2\right\}  $ (when Simes test is used) or $\hat{S}=\left\{ 1, 2,3\right\}$ (when the KS test is used) and the same
nominal FDR level $0.05$,
we applied the generic adaptive sGBH and generic adaptive GBH, and they identified the same number of differentially expressed genes. We caution that the FDRs of the adaptive GBH and sGBH
may exceed the specified normal level $0.05$ due to a potential violation of the assumptions
that ensure their non-asymptotic conservativeness.

\section{Discussion}

\label{secDis}

To better adapt to a group structure among hypotheses, we have proposed a
grouped, selectively weighted FDR procedure, sGBH, that is a refinement and
extension of the GBH and wFDR procedures of \cite{Hu:2010}, \cite{Chen:2015discretefdr} and \cite{Nandi:2018}
and that accommodates scenarios where only a few groups are likely to be interesting.
For the plug-in adaptive sGBH, we have provided simple conditions
to ensure and some empirical evidence on its conservativeness, together with an FDR upper bound
that quantifies the effect of estimating the interesting groups. Further,
we have provided numerical evidence on conservativeness and improved power of the generic adaptive sGBH that employs Simes test
to select interesting groups. These two versions of the adaptive sGBH have been numerically shown to be robust to the Type I error level
of the test that is used to select interesting groups and the tuning parameter that is used to construct the weights, and to be robust to
strongly-mixing dependence.
As with any grouped FDR procedure, how hypotheses are partitioned affects
inferential results, and we argue that this should usually be done carefully
using a practitioner's domain knowledge.

There are four issues left for future investigation. Firstly, it is worth developing an adaptive sGBH for multiple testing based on discrete p-values
that may utilize the consistent proportion estimators proposed by \cite{Chen:2018a} for discrete statistics.
Secondly, we have not numerically examined the relative performances
of the variant of sGBH and GBH since we were not able to identify configurations under
which the former is more powerful than the latter. However, we believe they do
exist. Thirdly, it is quite challenging to design a scheme to correctly,
non-asymptotically identify interesting and uninteresting groups under a nontrivial sparse configuration
 and show the non-asymptotic conservativeness of the resulting adaptive sGBH.
Fourthly, an unsettled issue is how to design data-adaptive weights that ensure the non-asymptotic conservativeness and improved power (with respect to BH)
of a weighted FDR procedure under dependence (or even under PRDS). For a proposal for this under block dependence, we refer the readers to \cite{Guo:2016}.

\bibliographystyle{dcu}


\appendix{}

\section{Proofs related to sGBH}

\label{secProofs}

In the proofs here and in the supplementary material, the indicator function $\mathbf{1}_{A}$ of a set $A$ will be written as $\mathbf{1}A$ if $A$ is
described by a proposition, and $\left\vert A\right\vert $ is the cardinality
of $A$.

\subsection{Proof of \autoref{ThmOracle}}

The first claim is obvious. Now we show the third claim. When $S=$
$\varnothing$, the oracle sGBH reduces to the oracle GBH with $\tilde{\pi}_0=\pi_{0}=1$,
makes no rejections and is thus conservative. So, it is left to consider the
case $S\neq$ $\varnothing$, which implies $1>\pi_{0}>\tilde{\pi}_{0}$. In this
case, the weights for p-values in the set $\mathbf{p}_{ S}=\left\{  p_{j_{k}%
}:j\in S,k\in G_{j}\right\}  $ are all finite, whereas $p_{j_{k}}$ is set to
be $\infty$ for each $j\notin S$ and $k\in G_{j}$. Therefore, the oracle sGBH
makes no false discoveries from the set $\mathcal{H}_{ S^{\prime}}=\left\{
H_{j_{k}}:j\notin S,k\in G_{j}\right\}  $, and we only need to study the
oracle GBH applied to $\mathbf{p}_{ S}$.

Even though Theorem 1 of \cite{Hu:2010} showed the conservativeness of the
oracle GBH when it is applied to $\mathbf{p}_{ S}$, missing is the
justification that the PRDS property of $\mathbf{p}_{ S}$ is preserved when
each $p_{i^{\prime}}\in\mathbf{p}_{ S}$ is weighted by a finite, nonnegative
deterministic number. Here we provide it. Recall $m_{ S}$ as the cardinality
of $\mathbf{p}_{ S}$, denote also by $\mathbf{p}_{ S}$ the vector formed by
enumerating elements of $\mathbf{p}_{ S}$, and let $R$ be the number of
rejections made by the oracle GBH when it is applied to the set $\tilde{\mathbf{p}}_{ S}$ of weighted p-values associated with the
interesting groups. Clearly, $R$ is
non-increasing in each p-value in $\mathbf{p}_{ S}$. In particular, for each
$r\in\left\{  0,\ldots m_{ S}\right\}  $,
\[
D_{r}=\left\{  \mathbf{p}_{ S}\in\left[  0,1\right]  ^{m_{ S}}:R<r\right\}
\]
is a non-decreasing set. Let $I_{0, S}$ be the index set of true nulls among
$\mathcal{H}_{ S}=\left\{  H_{j_{k}}:j\in S,k\in G_{j}\right\}  $. Then, the
PRDS property of $\mathbf{p}_{ S}$ and the arguments in the proof of
Proposition 3.6 of \cite{Blanchard:2008} imply that the function%
\begin{equation}
t\mapsto\Pr\left(  \left.  R<r\right\vert p_{i^{\prime}}\leq t\right)
\label{eqPRDS}%
\end{equation}
is nondecreasing for each $i^{\prime}\in I_{0, S}$ and $r\in\left\{  0,\ldots
m_{ S}\right\}  $. Thus, Theorem 1 of \cite{Hu:2010} yields the claim that the
oracle sGBH is conservative under PRDS. We remark that the conservativeness of the oracle sGBH
follows by slightly adapting the arguments of Theorem 1 of \cite{Nandi:2018} to super-uniform null p-values.

Finally, we show the second claim. Note that $m_{ S}$ is also the cardinality
of $\mathcal{H}_{ S}$. When $S\neq\varnothing$, we have%
\begin{equation}
1>\pi_{0}\geq\tilde{\pi}_{0}\text{ \ and \ }m\geq m_{ S}. \label{eq2a}%
\end{equation}
By the arguments presented above, to see which among the oracle GBH and the
oracle sGBH rejects more, we only need to check them based on $\tilde{\mathbf{p}}_{
S}$. Each $p_{j_{k}}\in\mathbf{p}_{ S}$ has been weighted into%
\begin{equation}
{p_{j_{k}}^{\ast}=}\frac{{p}_{j_{k}}\pi_{j0}}{1-\pi_{j0}}\left(  1-\pi
_{0}\right)  \text{ \ and \ }{\tilde{p}_{j_{k}}=}\frac{{p}_{j_{k}}\pi_{j0}%
}{1-\pi_{j0}}\left(  1-\tilde{\pi}_{0}\right)  \label{eq2b}%
\end{equation}
respectively by the oracle GBH and oracle sGBH. Let $G_{ S}=\left\{  j_{k}%
:{j}\in S\text{,}\ k\in G_{j}\right\}  $. Then (\ref{eq2a}), (\ref{eq2b}) and
no p-value taking value $0$ together imply ${p_{j_{k}}^{\ast}\leq\tilde
{p}_{j_{k}}}$ and%
\begin{equation}
\kappa=\frac{{\tilde{p}_{j_{k}}}}{{p_{j_{k}}^{\ast}}}=\frac{1-\tilde{\pi}_{0}%
}{1-\pi_{0}}\geq1\text{ for each }j_{k}\in G_{ S}. \label{eq2d}%
\end{equation}
Let the $i^{\prime}$-th order statistic among
$\left\{  \frac{{p}_{j_{k}}\pi_{j0}}{1-\pi_{j0}}:j_{k}\in G_{ S}\right\}  $ be
$\vartheta_{\left(  i^{\prime}\right)  }$. Then the $i^{\prime}$-th order statistic
among $\left\{  {p_{j_{k}}^{\ast}:}j_{k}\in G_{ S}\right\}  $, denoted by
${p_{\left(  i^{\prime}\right)  }^{\ast}}$, is $\vartheta_{\left(  i^{\prime
}\right)  }\left(  1-\pi_{0}\right)  $, and the $i^{\prime}$-th order statistic among
$\left\{  {\tilde{p}_{j_{k}}:}j_{k}\in G_{ S}\right\}  $, denoted by
${\tilde{p}_{\left(  i^{\prime}\right)  }}$, is $\vartheta_{\left(  i^{\prime
}\right)  }\left(  1-\tilde{\pi}_{0}\right)  $, and ${p_{\left(  i^{\prime
}\right)  }^{\ast}\leq\tilde{p}_{\left(  i^{\prime}\right)  }}$.

Let $R^{\ast}$ be the number of rejections made by the oracle GBH and
$\tilde{R}$ that by the oracle sGBH. Then $R^{\ast}\leq m_{S}$ and%
\[
{p_{\left(  R^{\ast}\right)  }^{\ast}\leq}\frac{R^{\ast}\alpha}{m}\text{ \ and
\ \ \ }{p_{\left(  r\right)  }^{\ast}}>\frac{r\alpha}{m}\text{ for all }%
m_{S}\geq r>R^{\ast}.
\]
With this, we see from (\ref{eq2d}), the identity
\begin{equation}
\frac{\left(  1-\tilde{\pi}_{0}\right)  m_{S}}{\left(  1-\pi_{0}\right)  m}=1
\label{eq2e}%
\end{equation}
and the orderings discussed in the previous paragraph that%
\[
{\tilde{p}_{\left(  R^{\ast}\right)  }\leq\kappa}\frac{R^{\ast}\alpha}%
{m}=\frac{\left(  1-\tilde{\pi}_{0}\right)  m_{S}}{\left(  1-\pi_{0}\right)
m}\frac{R^{\ast}\alpha}{m_{S}}=\frac{R^{\ast}\alpha}{m_{S}}%
\]
and ${\tilde{p}_{\left(  r\right)  }}>\kappa\frac{r\alpha}{m}=\frac{r\alpha
}{m_{S}}$ for $m_{S}\geq r>R^{\ast}$. On the other hand, $\tilde{R}\leq m_{S}$
and%
\[
\tilde{p}_{\left(  \tilde{R}\right)  }\leq\frac{\tilde{R}\alpha}{m_{S}}\text{
\ and \ \ \ }{\tilde{p}_{\left(  s\right)  }}>\frac{s\alpha}{m_{S}}\text{ for
all }m_{S}\geq s>\tilde{R}.
\]
So, from (\ref{eq2d}), (\ref{eq2e}) and the orderings discussed previously, we
see%
\[
p_{\left(  \tilde{R}\right)  }^{\ast}\leq\frac{1}{\kappa}\frac{\tilde{R}%
\alpha}{m_{S}}=\frac{\left(  1-\pi_{0}\right)  m}{\left(  1-\tilde{\pi}%
_{0}\right)  m_{S}}\frac{\tilde{R}\alpha}{m}=\frac{\tilde{R}\alpha}{m}%
\]
and $p_{\left(  s\right)  }^{\ast}>\frac{1}{\kappa}\frac{s\alpha}{m_{S}}%
=\frac{s\alpha}{m}$ for all $m_{S}\geq s>\tilde{R}$. Consequently, both
procedures reject the same set of hypotheses.

\subsection{Proof of \autoref{PropQgbh}}

\label{secProofQGBH}

Let $R^{\natural}$ and $\alpha_{m}^{\natural}$ be the number of rejections and
FDR\ of qGBH, respectively. Let $m_{S}=\sum\nolimits_{j\in S}n_{j}$,
$m_{\hat{S}}=\sum\nolimits_{j\in\hat{S}}n_{j}$ and%
\[
\pi_{0,\hat{S}}=\left(  \sum\nolimits_{j\in\hat{S}}n_{j}\pi_{j0}\right)
m_{\hat{S}}^{-1}.
\]
Note that each $\pi_{j0}$ is known to qGBH and $\pi_{j0}=1$ for $j\notin S$.
If $\hat{S}\cap S=\varnothing$, then $\pi_{0,\hat{S}}=1$, $R^{\natural}=0$ and
$\alpha_{m}^{\natural}=0$. On the other hand, if $\hat{S}\cap S\neq
\varnothing$ and $\hat{S}\cap\left(  \mathbb{N}_{l}\setminus S\right)
\neq\varnothing$, then $\pi_{0,\hat{S}}<1$ but no rejections will be made
from any group $\mathcal{H}_{j}$ for $j\notin S$. Without loss of generality, let $S=\left\{
1\right\}  $. Then $R^{\natural} \le n_1$ almost surely. Observing the identity $\left(1-\pi_{10}\right) n_1 = \left(1-\pi_{0,\hat{S}}\right) m_{\hat{S}}$
when $\hat{S}\cap S \ne \varnothing$, we have
\begin{align}
\alpha_{m}^{\natural}  &  \leq\mathbb{E}\left[  \sum_{j\in\hat{S}\cap S}%
\sum_{k\in S_{j0}}\frac{1}{R^{\natural}}\mathbf{1}\left\{  p_{j_{k}}\leq
\frac{1-\pi_{j0}}{\pi_{j0}}\frac{1}{1-\pi_{0,\hat{S}}}\frac{R^{\natural}%
\alpha}{m_{\hat{S}}}\right\}  \mathbf{1}_{\left\{  \hat{S}\cap S \ne \varnothing
\right\}  }\mathbf{1}_{\left\{ R^{\natural} \le n_1\right\} }\right] \nonumber\\
&  =\mathbb{E}\left[  \sum_{k\in S_{10}}\frac{1}{R^{\natural}}%
\mathbf{1}\left\{  p_{1_{k}}\leq\frac{1}{\pi_{10}}\frac{R^{\natural}\alpha
}{n_{1}}\right\} \mathbf{1}_{\left\{ R^{\natural} \le n_1\right\} } \right]  , \label{eqx4a}%
\end{align}
where $S_{j0}$ is the index set of true nulls in $\mathcal{H}_{j}$.
However, the quantity in (\ref{eqx4a}) is upper bounded by $\alpha$ by the
oracle property of sGBH under PRDS. Thus, the claim holds.

\subsection{Proof of \autoref{thmConserve}}

\label{SecProofAdaptive}

We will show the claims in two steps: \textbf{Step 1.} ``filter out irrelevant cases from the
analysis'' and \textbf{Step 2.} ``obtain upper bounds on $\hat{\alpha}_{m}$''.

\textbf{Step 1.} Recall $\mathbf{p}=\left(  p_{1},\ldots,p_{m}\right)  $ and
$R=R\left(  \mathbf{p}\right)  $ as the number of rejections made by the
procedure. Then, we can assume $R\left(  \mathbf{p}\right)  \geq1$. If
$\pi_{j0}=0$, then the null hypotheses in group $G_{j}$ are all false and do not
contribute to $\hat{\alpha}_{m}$. So, we can assume $\pi_{j0}>0$ for each $j$.
Since $\min\left\{  p_{i}:i\geq1\right\}  >0$ almost surely, a weighted
p-value is $0$ only when its associated weight is $0$. Further, it suffices to consider the
case $\hat{S} \ne \varnothing$.

\textbf{Step 2.} Before we proceed further, we need to set up some notations.
For each $j\in\mathbb{N}_{l}$, let $\mathbf{q}_{j}$ be the vector of p-values
whose indices are in group $G_{j}$, and for each $k\in G_{j}$, let $\mathbf{q}_{j,-k}$ be the
vector obtained by removing $p_{j_{^{k}}}$ from $\mathbf{q}_{j}$, and
$\mathbf{q}_{j,0,k}$ the vector obtained by setting $p_{j_{k}}=0$ in
$\mathbf{q}_{j}$. Let the plug-in adaptive sGBH employ the same proportion estimator
$\hat{\pi}_{0}^{\dagger}$ to estimate $\pi_{j0}$ for each needed
$j$. We will denote by $\hat{\pi}_{j0}$ and $\hat{\pi
}_{j0,-j_{k}}$ respectively the estimates obtained by applying $\hat{\pi}%
_{0}^{\dagger}$ to $\mathbf{q}_{j}$ and $\mathbf{q}_{j,0,k}$ for
a $j$ and $k\in G_{j}$. Recall $\hat{\pi}_{0} = m^{-1}\sum_{j=1}^l \hat{\pi}_{j0}n_j$. For any two vectors $\mathbf{\tilde{p}}$ and
$\mathbf{\hat{p}}$ whose entries together partition $\left\{  p_{1}%
,\ldots,p_{m}\right\}  $, we write $R\left(  \mathbf{p}\right)  $ equivalently
as $R\left(  \mathbf{p}\right)  =R\left(  \mathbf{\tilde{p}},\mathbf{\hat{p}%
}\right)  $. Let $V\left(  \mathbf{p}\right)  $ be the number of false
discoveries of the procedure, which is also written as $V$ for notational simplicity.
Let $m_{\hat{S}} = \sum_{j \in \hat{S}}n_j$ and
\[
\tilde{w}_{j}=\hat{\pi}_{j0}\left(  1-\hat{\pi}_{0,\hat{S}}\right)  \left(
1-\hat{\pi}_{j0}\right)  ^{-1}\text{ \ for each }j\in\hat{S},
\]
where we recall
\[
\hat{\pi}_{0,\hat{S}}=\left(  \sum\nolimits_{j\in\hat{S}}\hat{\pi}_{j0}%
n_{j}\right)  \left(  \sum\nolimits_{j\in\hat{S}}n_{j}\right)  ^{-1}.
\]

The rest of the proof will be divided
into 2 parts: \textbf{Part I} for the first claim and \textbf{Part II} the second.

\textbf{Part I:} Since $\Pr\left(  \hat{\pi}_{0,\hat{S}}<1\right)  >0$, there must be a
constant $\check{\pi}_{0}\in\lbrack0,1)$ such that the event $\mathcal{B}%
_{m}=\left\{  \hat{\pi}_{0,\hat{S}}\leq\check{\pi}_{0}\right\}  $ has positive
probability. Let $\mathcal{B}_{m}^{\prime}$ be the complement of
$\mathcal{B}_{m}$. Further, let $\mathcal{A}_{m}=\left\{  \hat{S}=S\right\}  $
with complement $\mathcal{A}_{m}^{\prime}$. Then%
\begin{equation}
\hat{\alpha}_{m}\leq\mathbb{E}\left[  \frac{V}{R}\mathbf{1}_{\mathcal{B}_{m}%
}\mathbf{1}_{\mathcal{A}_{m}}\right]  +\Pr\left(  \mathcal{B}_{m}^{\prime
}\right)  +\Pr\left(  \mathcal{A}_{m}^{\prime}\right)  . \label{eqx7e}%
\end{equation}
For each $j\in S$ and $k\in G_{j}$, let%
\begin{equation}
c_{j_{k}}=\frac{1-\hat{\pi}_{j0,-j_{k}}}{\hat{\pi}_{j0,-j_{k}}}\frac
{1}{\left(  1-\check{\pi}_{0}\right)  }. \label{eqx8a}%
\end{equation}
Since $\hat{\pi}_{0}^{\dagger}$ is a non-increasing estimator, we have almost
surely%
\[
\mathbf{1}\left\{  p_{j_{k}}\leq\frac{R\alpha}{\tilde{w}_{j}m_{\hat{S}}}\right\}
\mathbf{1}_{\left\{  j\in\hat{S}\right\}  }\mathbf{1}_{\mathcal{A}_{m}}\mathbf{1}_{\mathcal{B}_{m}}%
\leq\mathbf{1}\left\{  p_{j_{k}}\leq\frac{R\alpha}{c_{j_{k}}m_{S}}\right\}
\mathbf{1}_{\left\{  j\in S\right\}  }%
\]
for each $j\in\hat{S}$ and $k\in S_{j0}$. Therefore,%
\begin{equation}
\mathbb{E}\left[  \frac{V}{R}1_{\mathcal{B}_{m}}1_{\mathcal{A}_{m}}\right]
\leq\gamma_{m}=\sum_{r=1}^{m_{S}}\sum_{j\in S}\sum_{k\in S_{j0}}%
\vartheta_{j,k,r}, \label{eqx7f}%
\end{equation}
where for $j\in S$, $k\in S_{j0}$ and $1\leq r\leq m_{S}$,%
\[
\vartheta_{j,k,r}=\mathbb{E}\left[  \frac{1}{r}\mathbf{1}\left\{  p_{j_{k}%
}\leq\frac{r\alpha}{c_{j_{k}}m_{S}}\right\}  \mathbf{1}_{\left\{  R=r\right\}
}\right]
\]
However, when $\left\{ p_i \right\}_{i=1}^m$ are independent,
\begin{align*}
&  \sum_{r=1}^{m_{S}}\vartheta_{j,k,r}=\sum_{r=1}^{m_{S}}\mathbb{E}\left[
\mathbb{E}\left[  \left.  \frac{1}{r}\mathbf{1}\left\{  p_{j_{k}}\leq
\frac{r\alpha}{c_{j_{k}}m_{S}}\right\}  1_{\left\{  R=r\right\}  }\right\vert
\mathbf{p}_{-j_{k}}\right]  \right] \\
&  \leq \mathbb{E}\left[  \left.  \frac{\alpha
}{c_{j_{k}}m_{S}}\right\vert \mathbf{p}_{-j_{k}}\right]  =\frac{1}{\left(
1-\check{\pi}_{0}\right)  }\frac{\alpha}{m_{S}}\mathbb{E}\left[  \frac
{1-\hat{\pi}_{j0,-j_{k}}}{\hat{\pi}_{j0,-j_{k}}}\right] .
\end{align*}
Thus, $\gamma_{m}$ in (\ref{eqx7f}) satisfies%
\begin{align}
\gamma_{m}  &  \leq\frac{\alpha}{m_{S}}\frac{1}{\left(  1-\check{\pi}%
_{0}\right)  }\sum_{j\in S}\sum_{k\in S_{j0}}\mathbb{E}\left[  \frac
{1-\hat{\pi}_{j0,-j_{k}}}{\hat{\pi}_{j0,-j_{k}}}\right] \nonumber\\
&  \leq\frac{\alpha}{m_{S}}\frac{1}{\left(  1-\check{\pi}_{0}\right)  }%
\sum_{j\in S}\sum_{k\in S_{j0}}\frac{1-\pi_{j0}}{\pi_{j0}}=\frac{\alpha}%
{m_{S}\left(  1-\check{\pi}_{0}\right)  }\sum_{j\in S}n_{j}\left(  1-\pi
_{j0}\right)  , \label{eqx7h}%
\end{align}
where the inequality in (\ref{eqx7h}) holds since $\hat{\pi}_{0}^{\dagger}$
is reciprocally conservative, i.e., $\mathbb{E}\left[  \left.  1\right/
\hat{\pi}_{j0,-j_{k}}\right]  \leq\left.  1\right/  \pi_{j0}$. Combining
(\ref{eqx7e}), (\ref{eqx7f}) and (\ref{eqx7h}) gives (\ref{bndwFDR}), i.e.,%
\[
\hat{\alpha}_{m}\leq\frac{\alpha}{m_{S}}\frac{1}{1-\check{\pi}_{0}}%
\sum\nolimits_{j\in S}n_{j}\left(  1-\pi_{j0}\right)  +\Pr\left(  \hat{\pi
}_{0,\hat{S}}>\check{\pi}_{0}\right)  +\Pr\left(  \hat{S}\neq S\right)  .
\]

If in addition both $\hat{S}$ and $\hat{\pi}_{0,\hat{S}}$ are consistent, then
$\Pr\left(  \hat{\pi}_{0,\hat{S}}=\tilde{\pi}_{0}\right)
\rightarrow1$, $\Pr\left( \hat{S}=S \right)  \rightarrow1$ and the event $\mathcal{B}_{m}=\left\{  \hat{\pi}_{0,\hat{S}%
}\leq\check{\pi}_{0}\right\}  $ can be set as $\left\{
\hat{\pi}_{0,\hat{S}}\leq\tilde{\pi}_{0}\right\}  $. By almost identical
arguments used to obtain (\ref{bndwFDR}), we have
\begin{align*}
\hat{\alpha}_{m}  &  \leq\frac{\alpha}{m_{S}}\frac{1}{1-\tilde{\pi}_{0}}%
\sum\nolimits_{j\in S}n_{j}\left(  1-\pi_{j0}\right)  +\Pr\left(  \hat{\pi
}_{0,\hat{S}}>\tilde{\pi}_{0}\right)  +\Pr\left(  \hat{S}\neq S\right) \\
&  \leq\alpha+\Pr\left(  \hat{\pi}_{0,\hat{S}}>\check{\pi}_{0}\right)
+\Pr\left(  \hat{S}\neq S\right)  .
\end{align*}
Therefore, $\limsup_{m\rightarrow\infty}\hat{\alpha}_{m}\leq\alpha$.

\textbf{Part II}: To show $\limsup_{m\rightarrow\infty}\hat{\alpha}_{m}%
\leq\alpha$ when $\left\{p_{i}\right\}_{i=1}^{m}$ have the property of PRDS, $\Pr(\hat{S}\neq S)\rightarrow0$ and $\Pr\left(  \bigcap\nolimits_{j\in \hat{S}}\left\{  \hat{\pi}%
_{j0}=\pi_{j0}\right\}  \right)  \rightarrow1$. Let $\mathcal{C}_{m}=\bigcap\nolimits_{j\in S}\left\{
\hat{\pi}_{j0}=\pi_{j0}\right\}  $ and $\mathcal{D}_{m}$ be the complement of
$\mathcal{C}_{m}$. Then,%
\begin{align*}
\hat{\alpha}_{m}  &  \leq\sum_{j\in S}\sum_{k\in S_{j0}}\mathbb{E}\left[
\frac{1}{R}\mathbf{1}\left\{  p_{j_{k}}\frac{\pi_{j0}\left(  1-\tilde{\pi
}_0\right)  }{1-\pi_{j0}}\leq\frac{R\alpha}{m}\right\}  \right]  +\Pr\left(
\mathcal{D}_{m}\right)  +\Pr\left(  \hat{S}\neq S\right) \\
&  \leq\alpha+\Pr\left(  \mathcal{D}_{m}\right)  +\Pr\left(  \hat{S}\neq
S\right)  ,
\end{align*}
where the second inequality follows from the conservativeness of the oracle sGBH under
PRDS. So, the claim holds.

\subsection{Proof of \autoref{ThmEstS}}

For each $j\in\left\{  1,\ldots,l\right\}  $, let $F_{j}$ be the empirical
distribution of the p-values whose indices are in group $G_{j}$. Let $m_{j0}=n_{j}\pi_{j0}$ and
$m_{j1}=n_{j}\left(  1-\pi_{j0}\right)  $. First of all, for each $j$,%
\[
F_{j}\left(  t\right)  -t=\pi_{j0}d_{j,0,m}+\left(  1-\pi_{j0}\right)  d_{j,1,m},
\]
where $d_{j,0,m}\left(  t\right)  ={m_{j0}^{-1}}\sum\nolimits_{i\in S_{j0}%
}\left(  \mathbf{1}_{\left\{  p_{i}\leq t\right\}  }-t\right)  $ and%
\[
d_{j,1,m}\left(  t\right)  =\frac{1}{m_{j1}}\sum\nolimits_{i\in G_{j}\setminus
S_{j0}}\left(  \mathbf{1}_{\left\{  p_{i}\leq t\right\}  }-t\right)  .
\]
By the independence between $\left\{  p_{i}\right\}  _{i=1}^{m}$, we see that
$\sup_{t\in\left[  0,1\right]  }\left\vert d_{j,0,m}\left(  t\right)
\right\vert \rightarrow$ $0$ almost surely uniformly in $j\notin S$. By
assumption (\ref{condEstS}) on $d_{j,1,m}\left(  t\right)  $, we see that%
\begin{equation}
\lim_{m\rightarrow\infty}\inf\nolimits_{j\in S}\sqrt{n_{j}}\sup\nolimits_{t\in
\left[  0,1\right]  }\left\vert F_{j}\left(  t\right)  -t\right\vert
=\infty\text{.} \label{eqx15b}%
\end{equation}
By \cite{massey1950}, (\ref{eqx15b}) implies that the power of the KS test
tends to $1$ as $m\rightarrow\infty$ uniformly in $j\in S$. Therefore, $\Pr(
\hat{S}= S)\rightarrow1$.

\newpage
\renewcommand*{\thefootnote}{\fnsymbol{footnote}}
\begin{center}
\title{{\Large{Supplementary material for ``A grouped, selectively weighted false discovery rate procedure"\\
}}}
\bigskip
\author{Xiongzhi Chen\footnote{Corresponding author: Department of Mathematics and
Statistics, Washington State University, Pullman, WA 99164, USA; Email:
\texttt{xiongzhi.chen@wsu.edu}.} \ and Sanat K.
Sarkar\footnote{Department of Statistical Science and Fox School of Business,
Temple University, Philadelphia, PA 19122, USA; Email:
\texttt{sanat@temple.edu}.} }
\end{center}

We provide 
in \autoref{secVariant} a variant of sGBH
and its properties, and in \autoref{secAddSim} additional simulation results.
The powers of the plug-in adaptive sGBH and plug-in adaptive GBH based on Storey's estimator and one-sided p-values are
very close to zero and hence not reported.

\renewcommand\thefigure{\thesection.\arabic{figure}} \setcounter{figure}{0}

\section{A variant of sGBH}

\label{secVariant}

In this section, we introduce a variant of the oracle sGBH outside the setting
of sparse configuration. The notations in this section bear the same meanings as those for sGBH unless otherwise noted or
defined. The main message is that, for non-sparse configurations, the variant
can be more powerful than the GBH, whereas for sparse configurations it cannot.

Let $S$ be a subset of $\mathbb{N}${$_{l}$}, i.e., $S$ is the set of interesting groups of hypotheses, for which $\pi
_{j0}=1$ is not necessarily required for any $j\notin S$. Set the weights as%
\begin{equation}
w_{j}=\left\{
\begin{array}
[c]{ccc}%
\frac{\pi_{j0}\left(  1-\tilde{\pi}_{0}\right)  }{1-\pi_{j0}} & \text{if} &
j\in S\\
1 & \text{if} & j\notin S
\end{array}
\right.  , \label{eqe8a}%
\end{equation}
where $w_{j}=\infty$ is set when $\tilde{\pi}_{0}=1$ and/or $\pi_{j0}=1$.
{Weight each p-values $p_{i}$ into $\tilde{p}_{i}=p_{i}w_{j}$ for }$i\in
G_{j}$ for each $j\in\mathbb{N}_{l}${. Apply the BH procedure to the }$m$
weighted p-values $\left\{  {\tilde{p}_{i}}\right\}  _{i=1}^{m}$. Unless
otherwise noted, we will refer to the above procedure as the
``variant''. Note that $S$ is preselected and do not have to be
estimated, and that only p-values in the preselected groups are effectively weighted.

We assume that the same hypotheses configuration is used for both the
oracle GBH and the variant and that the same proportion estimator is employed
by the plug-in adaptive version of the variant (``adaptive variant'' for short) and the plug-in adaptive GBH
to obtain the plug-in weights.

\begin{theorem}
\label{ThmOracleA}If $S=\mathbb{N}${$_{l}$, }then the variant coincides with
the oracle GBH. However, under a nontrivial sparse configuration,
the variant never rejects more false nulls than the oracle GBH at the same
nominal FDR level. When $\left\{  p_{i}\right\}  _{i=1}^{m}$ have the property of
PRDS and $\tilde{\pi}_0 <1$, the variant is conservative.
\end{theorem}

\autoref{ThmOracleA} implies that one should not attempt a method as the
variant under a nontrivial sparse configuration in order to reject more false
nulls than the oracle GBH. On the other hand, one may attempt to identify
conditions under which the uniform dominance%
\begin{equation}
p_{j_{k}}^{\ast}\geq \text{\ }{\tilde{p}_{j_{k}}}\text{ \ for }j\in
\mathbb{N}_{l}\text{ and }k\in G_{j} \label{eq1x4}%
\end{equation}
holds and thus the variant rejects no less hypotheses than the oracle GBH,
where $p_{j_{k}}^{\ast}=p_{j_{k}}v_j$ and $\tilde{p}_{j_{k}}=p_{j_{k}}w_j$ for $j\in \mathbb{N}_{l}$ and $k\in G_{j}$. We will
present two scenarios where (\ref{eq1x4}) can never hold. Set the proportion
of true nulls for the uninteresting groups as
\begin{equation}
\rho_{0}=\left(  \sum\nolimits_{j\notin S}n_{j}\right)  ^{-1}\sum
\nolimits_{j\notin S}\pi_{j0}n_{j} \label{eqe8c}%
\end{equation}
and let%
\begin{equation}
\varsigma=\min_{j\notin S}\frac{\pi_{j0}\left(  1-\pi_{0}\right)  }{1-\pi
_{j0}}. \label{eqe8}%
\end{equation}

\begin{proposition}
\label{ThmImprovementA}Assume $S\neq\varnothing$ and $S\neq\mathbb{N}_{l}$. If
$\max\left\{  \pi_{0},\tilde{\pi}_{0}\right\}  <1$, then it cannot hold that
$\varsigma\geq1$ and $\tilde{\pi}_{0}\geq\rho_{0}$. On the other hand, if
$\tilde{\pi}_{0}=1$ but $\pi_{0}<1$, then $\varsigma\geq1$ cannot hold.
\end{proposition}

In \autoref{ThmImprovementA}, the opposite of either claim under its
corresponding settings implies (\ref{eq1x4}). We remark that the pessimistic
conclusion from \autoref{ThmImprovementA} does not mean that there is no other
setting where the variant is more powerful than the oracle GBH. However, we
find it very challenging to identify such settings. Recall
\[
\hat{\pi}_{0,S}=\left(  \sum\nolimits_{j\in S}\hat{\pi}_{j0}n_{j}\right)
\left(  \sum\nolimits_{j\in S}n_{j}\right)  ^{-1}%
\]
and let $\tilde{\alpha}_{m}$ be the FDR of the adaptive variant. We have

\begin{theorem}
\label{thmConserveA}Assume $\tilde{\pi}_{0}\in\lbrack0,1)$ uniformly in $m$
and that $\left\{  p_{i}\right\}  _{i=1}^{m}$ are mutually independent. If each $\hat{\pi}_{j0}, j \in S$
is non-increasing and reciprocally conservative and
$\Pr\left(  \hat{\pi}_{0,S}<1\right)  >0$, then there exits a constant
$\check{\pi}_{0}\in\lbrack0,1)$ such that $\Pr\left(  \hat{\pi}_{0,S}%
\leq\check{\pi}_{0}\right)  >0$ and
\begin{equation}
\tilde{\alpha}_{m}\leq\frac{\alpha}{m}\frac{1}{1-\check{\pi}_{0}}%
\sum\nolimits_{j\in S}n_{j}\left(  1-\pi_{j0}\right)  +\frac{\alpha}{m}%
\sum\nolimits_{j\notin S}n_{j}\pi_{j0}+\Pr\left(  \hat{\pi}_{0,S}>\check{\pi
}_{0}\right)  . \label{bndwFDRa}%
\end{equation}
If in addition $\hat{\pi}_{0,S}$ consistently estimates $\tilde{\pi}_0$, then $\limsup_{m\rightarrow
\infty}\tilde{\alpha}_{m}\leq\alpha$. On the other hand, if $\left\{p_{i}\right\}_{i=1}^{m}$ have the property of PRDS and $\hat{\pi}_{j0}$
is consistent for $\pi_{j0}$ uniformly in $j\in S$ (with necessarily being non-increasing or reciprocally conservative), then $\limsup_{m\rightarrow\infty}%
\hat{\alpha}_{m}\leq\alpha$.
\end{theorem}

\autoref{thmConserveA} bears the same spirit as \autoref{thmConserve} and
generalizes Theorem 3 of \cite{Chen:2015discretefdr}.
The (adaptive) variant allows a user to select groups of hypotheses of
interest and is very flexible.

\subsection{Proof of \autoref{ThmOracleA}}

The first claim is obvious. Now we show the third claim. Let $R$ be the
number of rejections made by the variant. From the proof of
\autoref{ThmOracle}, we see that
\begin{equation}
t\mapsto\Pr\left(  \left.  R<r\right\vert p_{i}\leq t\right)  \label{eqe5a}%
\end{equation}
is nondecreasing for each $i\in I_{0}$ and $r\in\left\{  0,\ldots m\right\}
$. Recall $w_{j}=\frac{\pi_{j0}\left(  1-\tilde{\pi}_{0}\right)  }{1-\pi_{j0}%
}$ for $j\in S$ and $w_{j}=1$ for $j\notin S$. Since $\tilde{\pi}_{0}<1$, each
weight $w_{j},j=1,\ldots,l$ is positive and finite. Let $\hat{\alpha}$ be the
FDR of the variant. Then,%
\begin{equation}
\hat{\alpha}\leq\frac{\alpha}{m}\sum_{k\in S_{j0}}\sum_{j=1}^{l}\frac{1}%
{w_{j}}\sum_{r=1}^{m}\Pr\left(  \left.  R=r\right\vert p_{j_{k}}\leq
\frac{r\alpha}{w_{j}m}\right)  . \label{eqex}%
\end{equation}
From (\ref{eqe5a}), we obtain, for each fixed $j\in\left\{  1,\ldots
,l\right\}  $ and $k\in G_{j}$,%
\begin{align}
&  \sum_{r=1}^{m}\Pr\left(  \left.  R=r\right\vert p_{j_{k}}\leq\frac{r\alpha
}{w_{j}m}\right) \nonumber\\
&  \leq\sum_{r=1}^{m}\left[  \Pr\left(  \left.  R\geq r\right\vert p_{j_{k}%
}\leq\frac{r\alpha}{w_{j}m}\right)  -\Pr\left(  \left.  R\geq r+1\right\vert
p_{j_{k}}\leq\frac{\left(  r+1\right)  \alpha}{w_{j}m}\right)  \right]
\nonumber\\
&  =1. \label{eqe6a}%
\end{align}
Thus, (\ref{eqex}) and (\ref{eqe6a}) together imply $\hat{\alpha}\leq
\frac{\alpha}{m}\sum_{k\in S_{j0}}\sum_{j=1}^{l}\frac{1}{w_{j}}$. However,%
\[
\frac{1}{m}\sum_{k\in S_{j0}}\sum_{j=1}^{l}\frac{1}{w_{j}}=\frac{1}{m}%
\sum\nolimits_{j\in S}n_{j}+\frac{1}{m}\sum\nolimits_{j\notin S}n_{j}\pi
_{j0}\leq1.
\]
So, $\hat{\alpha}\leq\alpha$, and the variant is conservative.

Finally, we show the second claim. Under the assumptions, the weights
$v_{j}=\infty$ for $j\notin S$, i.e., ${p_{j_{k}}^{\ast}=\infty}$ for $j\notin
S$ and $k\in G_{j}$, the weights $w_{j}$ and $v_{j}$ for $j\in S$ are all
positive and finite, and both $\pi_{0}$ and $\tilde{\pi}_{0}$ are less than $1$. It is natural to look at the relative orders between%
\begin{equation}
p_{j_{k}}^{\ast}=p_{j_{k}}\frac{\pi_{j0}\left(  1-\pi_{0}\right)  }{1-\pi_{j0}}\text{
\ and \ }\tilde{p}_{j_{k}}=p_{j_{k}}\frac{\pi_{j0}\left(  1-\tilde{\pi}_{0}\right)
}{1-\pi_{j0}} \label{eqx3g}%
\end{equation}
for ${j}\in S$ and $k\in G_{j}$ and $p_{j_{k^{\prime}}^{\prime}}$ for
${j}^{\prime}\notin S$ and $k^{\prime}\in G_{j^{\prime}}$. Under a nontrivial
sparse configuration, $\rho_{0}=1$ and the identity
$\pi_{0}-\tilde{\pi}_{0}=m^{-1}\left(  1-\tilde{\pi}_{0}\right)\sum\nolimits_{j\notin S}n_{j}$
implies $\pi_{0}\geq\tilde{\pi}_{0}$. This, together with (\ref{eqx3g}),
implies ${p_{j_{k}}^{\ast}\leq\tilde{p}_{j_{k}}}$ for all\ ${j}\in S$
and\ $k\in G_{j}$. So, the variant can never reject more false nulls than the
oracle GBH, and the former may reject more hypotheses than the latter only by
rejecting some true nulls.

\subsection{Proof of \autoref{ThmImprovementA}}

Let $s_{0}=\sum\nolimits_{j\in S}n_{j}$ and $q_{0}=\sum\nolimits_{j\notin
S}n_{j}$. Consider the first claim. Let $a=\left(  1-\pi_{0}\right)  ^{-1}$.
Then $\frac{a}{1+a}=\frac{1}{2-\pi_{0}}$. Suppose $\varsigma\geq1$ and
$\tilde{\pi}_{0}\geq\rho_{0}$. Then $\min_{j\notin S}\pi_{j0}\geq\frac
{1}{2-\pi_{0}}$ and $\rho_{0}\geq\frac{1}{2-\pi_{0}}$. When $\tilde{\pi}%
_{0}\geq\rho_{0}$, we must have $\rho_{0}\leq\pi_{0}\leq\tilde{\pi}_{0}$. In
other words, (\ref{eqe8}) implies $\pi_{0}\geq\frac{1}{2-\pi_{0}}$. However,
this forces $\pi_{0}=1$, contradicting $\max\left\{  \pi_{0},\tilde{\pi}%
_{0}\right\}  <1$. Now consider the second claim. Suppose $\varsigma\geq1$.
Then%
\begin{equation}
\pi_{0}^{2}\left(  s_{0}+q_{0}\right)  -\left(  3s_{0}+q_{0}\right)  \pi
_{0}+2s_{0}+q_{0}\leq0 \label{eq1x3}%
\end{equation}
has to hold. This forces $\pi_{0}\in\left[  1,2\right]  $, contradicting
$\pi_{0}<1$.

\subsection{Proof of \autoref{thmConserveA}}

\label{SecProofAdaptiveA}

The arguments to be presented next are very similar to those in the proof of
\autoref{thmConserve}, and the notations here bear the same meanings there
unless otherwise defined or noted. It suffices to consider the setting where the number of rejections of the procedure
$R=R\left(  \mathbf{p}\right)  \geq1$ and $\pi_{j0}>0$ for each $j$. The rest of
the proof will be divided into 2 parts: \textbf{Part I} for the first claim and \textbf{Part II} the second.

\textbf{Part I:} Let $\check{w}_{j}=c_{j_{k}}$ for $j\in S$ and $\check{w}_{j}=1$ for $j\notin
S$, where $c_{j_{k}}$ is defined by (\ref{eqx8a}). For each $j\in S$ and $k\in
S_{j0}$, define%
\[
\theta_{j_{k}}=\mathbb{E}\left[  \left.  \frac{1}{R\left(  p_{j_{k}%
},\mathbf{p}_{-j_{k}}\right)  }\mathbf{1}\left\{  p_{j_{k}}\leq\frac{R\left(
p_{j_{k}},\mathbf{p}_{-j_{k}}\right)  \alpha}{\check{w}_{j}m}\right\}
\right\vert \mathbf{p}_{-j_{k}}\right]  .
\]
The same strategy in \textbf{Part I} of the proof of \autoref{thmConserve}
implies
\[
\tilde{\alpha}_{m}\leq\sum\nolimits_{j=1}^{l}\sum\nolimits_{k\in S_{j0}%
}\mathbb{E}\left[  \theta_{j_{k}}\right]  +\Pr\left(  \hat{\pi}_{0}>\check
{\pi}_{0}\right)
\]
and%
\[
\sum_{j=1}^{l}\sum_{k\in S_{j0}}\mathbb{E}\left[  \theta_{j_{k}}\right]
\leq\frac{\alpha}{m\left(  1-\check{\pi}_{0}\right)  }\sum_{j\in S}%
n_{j}\left(  1-\pi_{j0}\right)  +\frac{\alpha}{m}\sum\nolimits_{j\notin
S}n_{j}\pi_{j0}\text{.}%
\]
Therefore, we have (\ref{bndwFDRa}), i.e.,%
\[
\tilde{\alpha}_{m}\leq\frac{\alpha}{m}\frac{1}{1-\check{\pi}_{0}}%
\sum\nolimits_{j\in S}n_{j}\left(  1-\pi_{j0}\right)  +\frac{\alpha}{m}%
\sum\nolimits_{j\notin S}n_{j}\pi_{j0}+\Pr\left(  \hat{\pi}_{0}>\check{\pi
}_{0}\right)  .
\]

If in addition $\Pr\left(  \hat{\pi}_{0,S}=\tilde{\pi}_{0}\right)
\rightarrow1$, then the same strategy in \textbf{Part I} of the proof of
\autoref{thmConserve} gives%
\begin{align*}
\tilde{\alpha}_{m}  &  \leq\frac{\alpha}{m}\frac{1}{1-\tilde{\pi}_{0}}%
\sum\nolimits_{j\in S}n_{j}\left(  1-\pi_{j0}\right)  +\frac{\alpha}{m}%
\sum\nolimits_{j\notin S}n_{j}\pi_{j0}+\Pr\left(  \hat{\pi}_{0,S}>\tilde{\pi
}_{0}\right) \\
&  \leq\alpha+\Pr\left(  \hat{\pi}_{0,S}>\tilde{\pi}_{0}\right)  ,
\end{align*}
which yields $\limsup_{m\rightarrow\infty}\hat{\alpha}_{m}\leq\alpha$.

\textbf{Part II}: To show $\limsup_{m\rightarrow\infty}\hat{\alpha}_{m}%
\leq\alpha$ when $\left\{p_{i}\right\}_{i=1}^{m}$ have the property of PRDS
and $\Pr\left(  \bigcap\nolimits_{j\in S}\left\{  \hat{\pi}%
_{j0}=\pi_{j0}\right\}  \right)  \rightarrow1$. Let $\mathcal{D}_{m}$ be
the complement of the event $\bigcap\nolimits_{j\in S}\left\{  \hat{\pi}%
_{j0}=\pi_{j0}\right\}  $. Then
\begin{align*}
\hat{\alpha}_{m}  &  \leq\sum_{j=1}^{l}\sum_{k\in S_{j0}}\mathbb{E}\left[
\frac{1}{R}\mathbf{1}\left\{  p_{j_{k}}w_{j}\leq\frac{R\alpha}{m}\right\}
\right]  +\Pr\left(  \mathcal{D}_{m}\right)   \leq\alpha+\Pr\left(  \mathcal{D}_{m}\right)  ,
\end{align*}
where the second inequality follows from the conservativeness of the variant under PRDS. So, the
claim holds.

\clearpage
\section{Additional simulation results}

\label{secAddSim}

\begin{figure}[H]
\centering
\includegraphics[height=0.79\textheight,width=\textwidth]{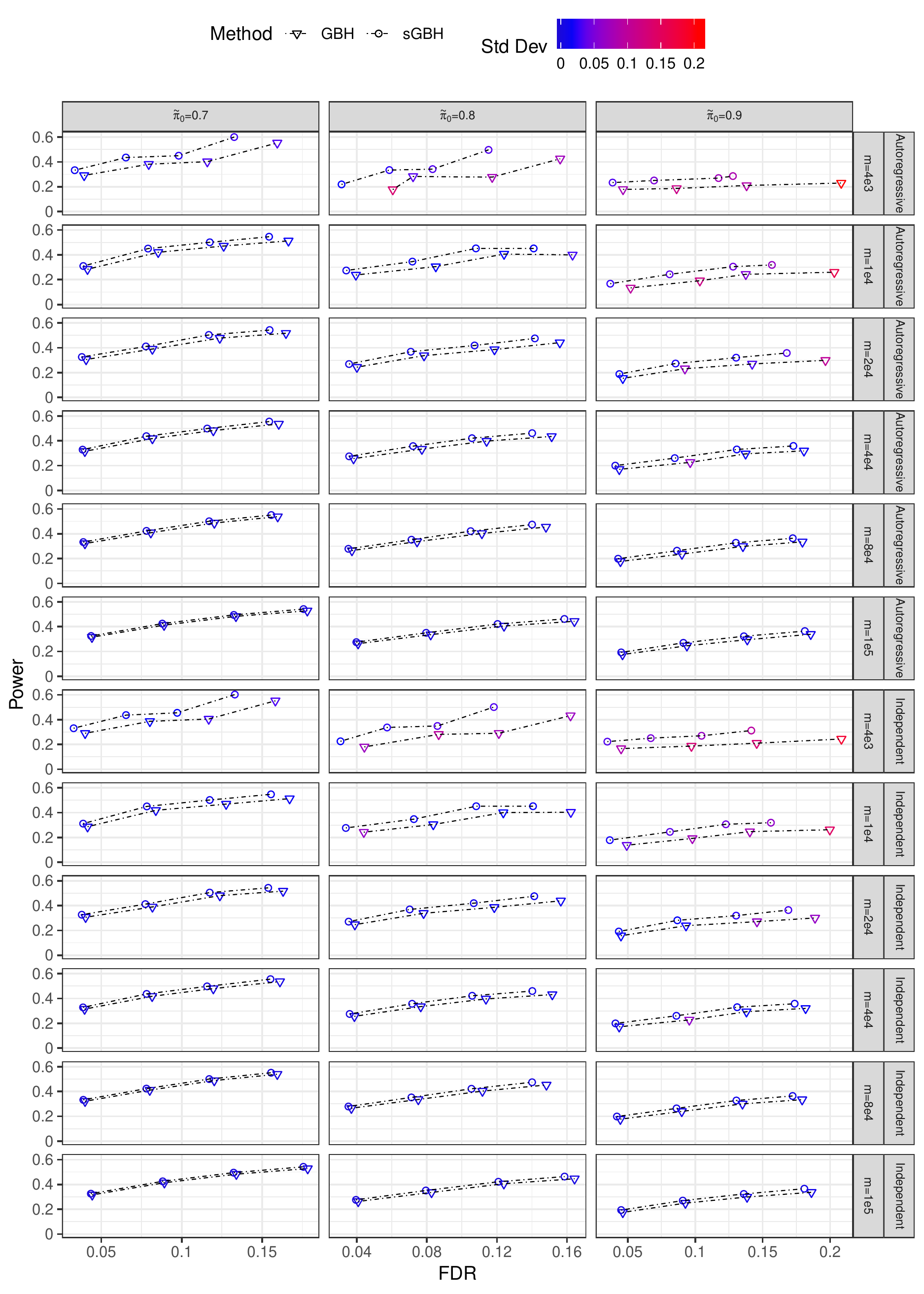}
\vspace{-0.5cm}\caption[power independence two-sided, storey est]{FDRs and powers
of the plug-in adaptive sGBH (``sGBH'') and GBH (``GBH'') based on two-sided p-values. Each type of points from left to
right in each subfigure are obtained successively under nominal FDR level
$0.05,0.1,0.15$ and $0.2$. The color legend ``Std Dev'' is the standard deviation of the FDP. The KS test has been used to identify interesting groups, and Storey's estimator to estimate the null
proportions.}%
\label{figPowerStorey2Side}%
\end{figure}

\begin{figure}[H]
\centering
\includegraphics[height=0.85\textheight,width=\textwidth]{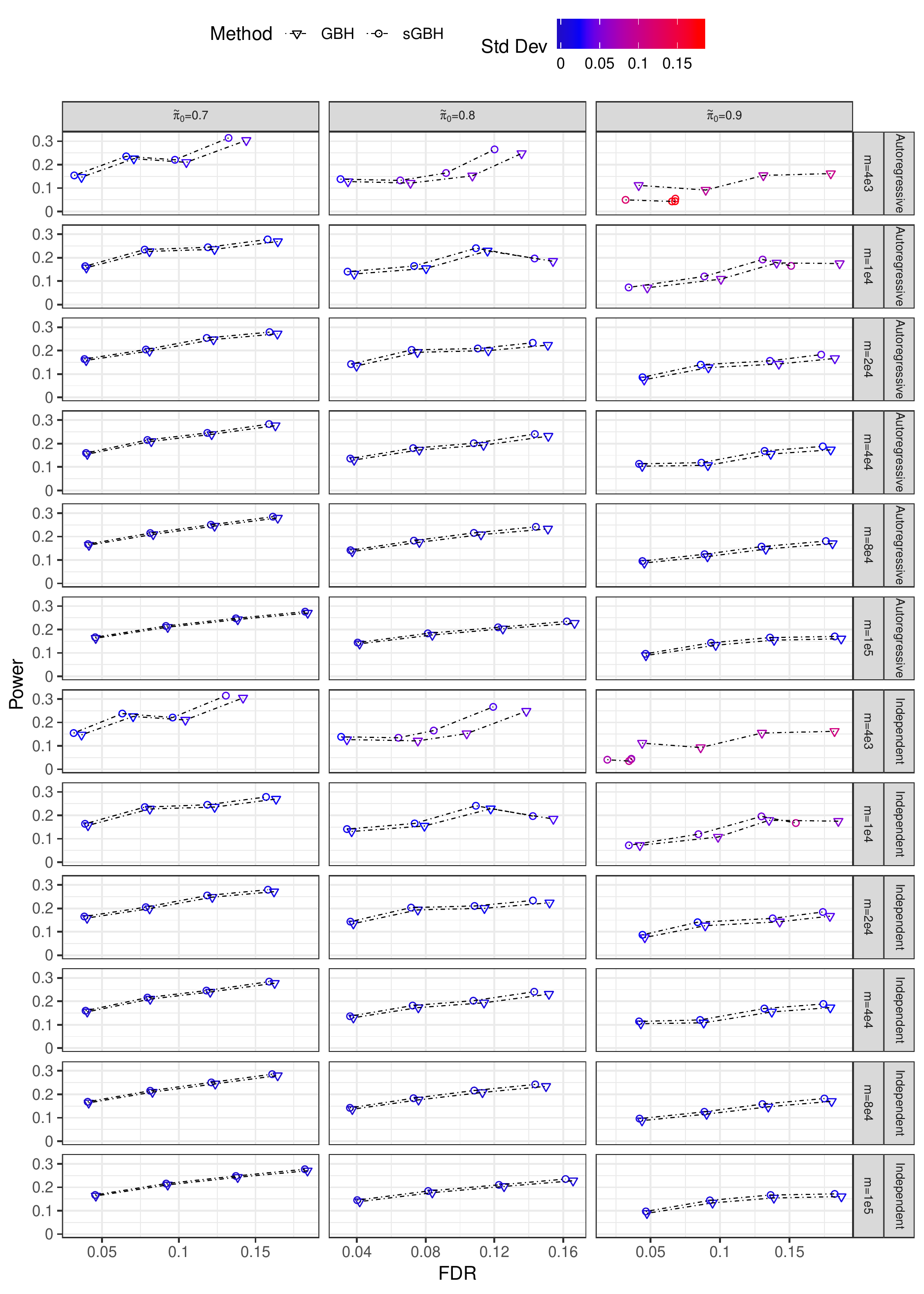}
\vspace{-0.5cm}\caption[power independence 1-sided jin est]{FDRs and powers of
the plug-in adaptive sGBH (``sGBH'') and GBH (``GBH'') based on one-sided p-values. Each type of points from left to
right in each subfigure are obtained successively under nominal FDR level
$0.05,0.1,0.15$ and $0.2$. The color legend ``Std Dev'' is the standard deviation of the FDP. Jin's estimator has been used to estimate the null
proportions, and the KS test to identify interesting groups.}%
\label{figPowerJin1Side}%
\end{figure}

\begin{figure}[H]
\centering
\includegraphics[height=0.85\textheight,width=0.95\textwidth]{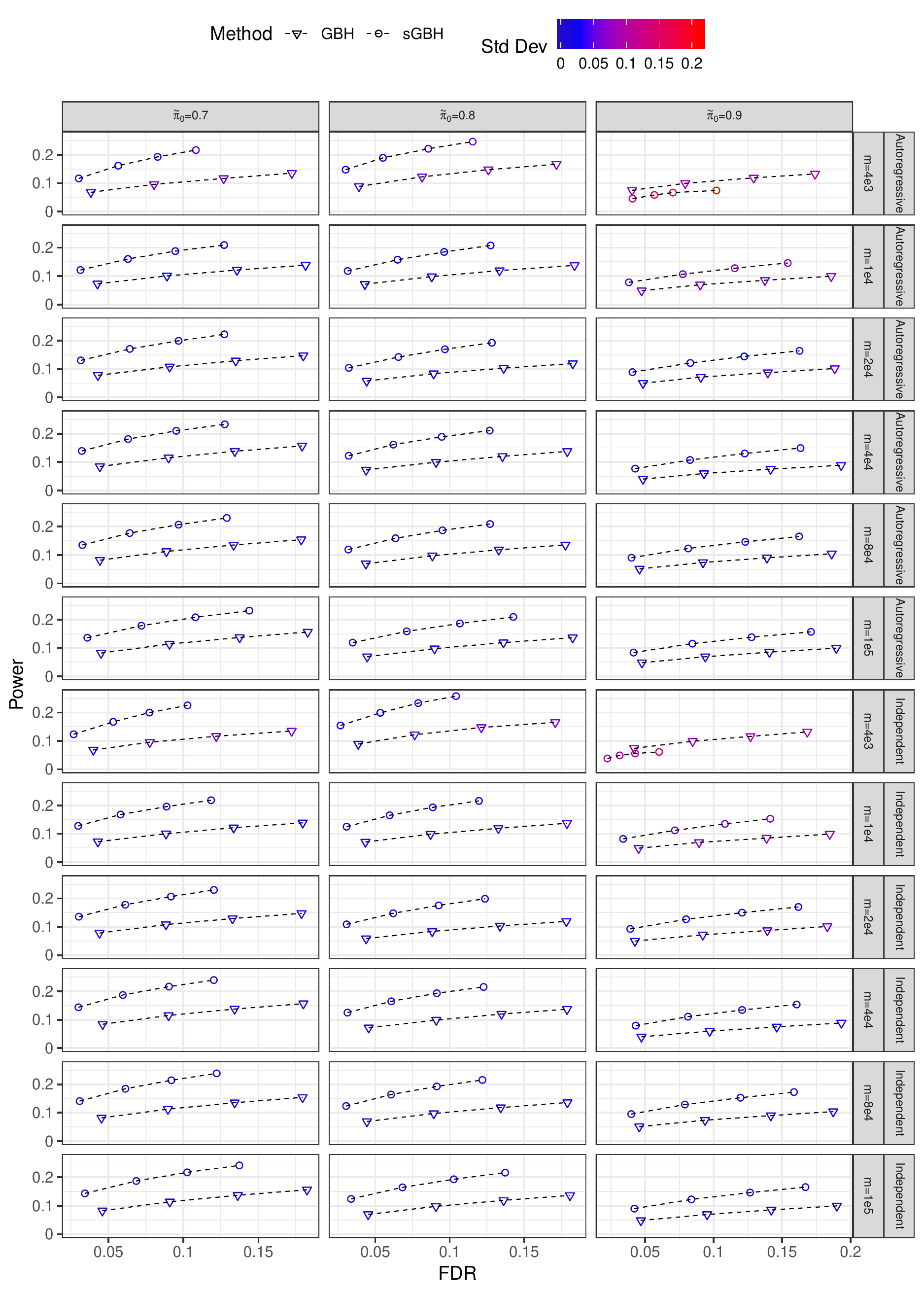}
\vspace{-0.5cm}\caption[power new, 1-side]{FDRs and powers of
the generic adaptive sGBH (``sGBH'') and GBH (``GBH'') based on one-sided p-values and tuning parameter $\lambda=0.5$. Each type of points from left to
right in each subfigure are obtained successively under nominal FDR level
$0.05,0.1,0.15$ and $0.2$. The color legend ``Std Dev'' is the standard deviation of the FDP. The KS test has been used to identify interesting groups.}%
\label{figPowerNew1Side}%
\end{figure}

\begin{figure}[H]
\centering
\includegraphics[height=0.86\textheight,width=.95\textwidth]{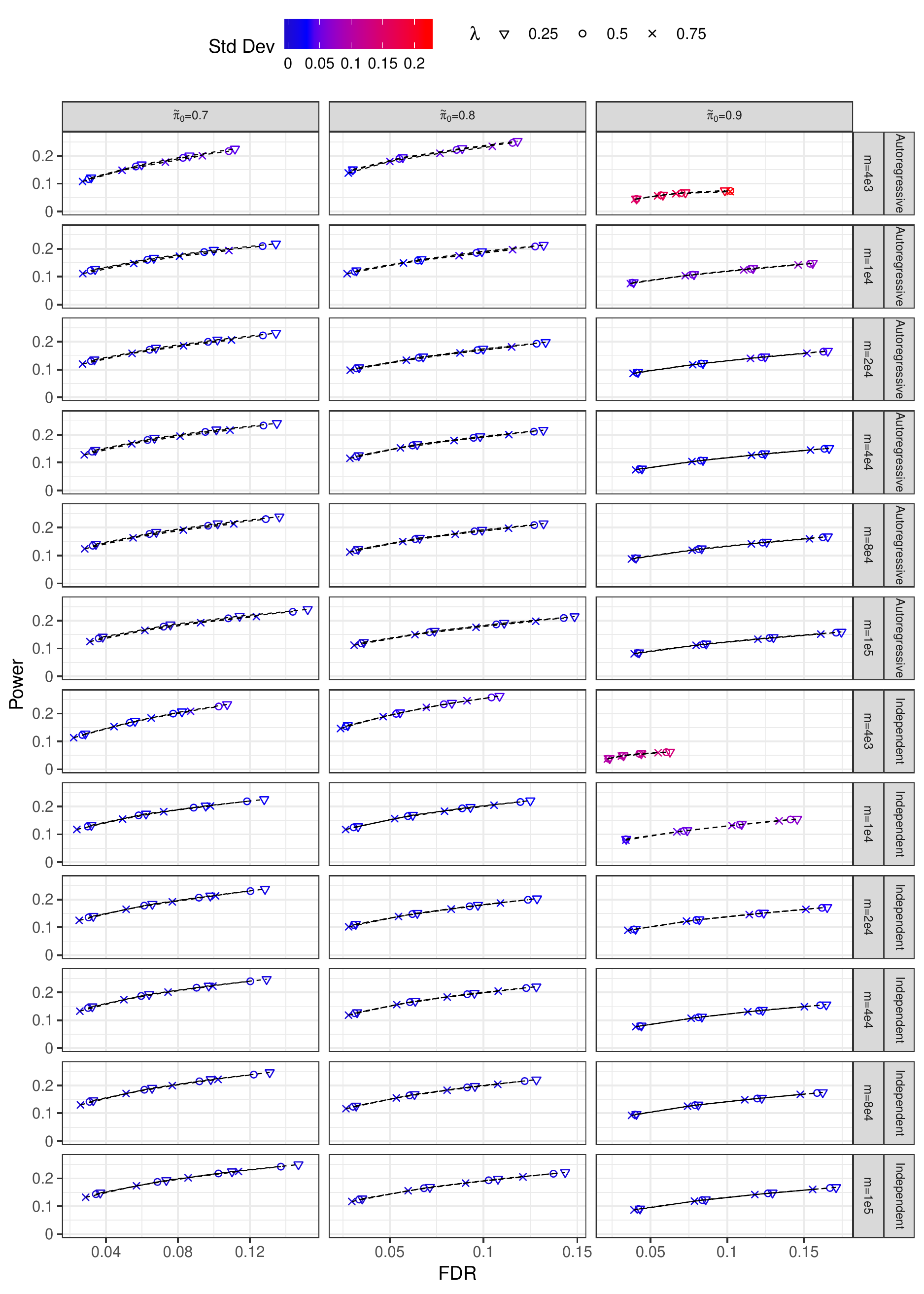}
\vspace{-0.5cm}\caption[power new, 1-sided,lambda]{FDR and power of the generic
adaptive sGBH (``sGBH") based on one-sided p-values and as the tuning
parameter $\lambda$ (shown in the legend) ranges in $\{0.25,0.5,0.7\}$. Each
type of points from left to right in each subfigure are obtained successively
under nominal FDR level $0.05,0.1,0.15$ and $0.2$. The color legend ``Std Dev'' is the standard deviation of the FDP. The KS test has been used to identify interesting groups.}%
\label{figPowerNewLambda1side}%
\end{figure}

\begin{figure}[H]
\centering
\includegraphics[height=0.87\textheight,width=.95\textwidth]{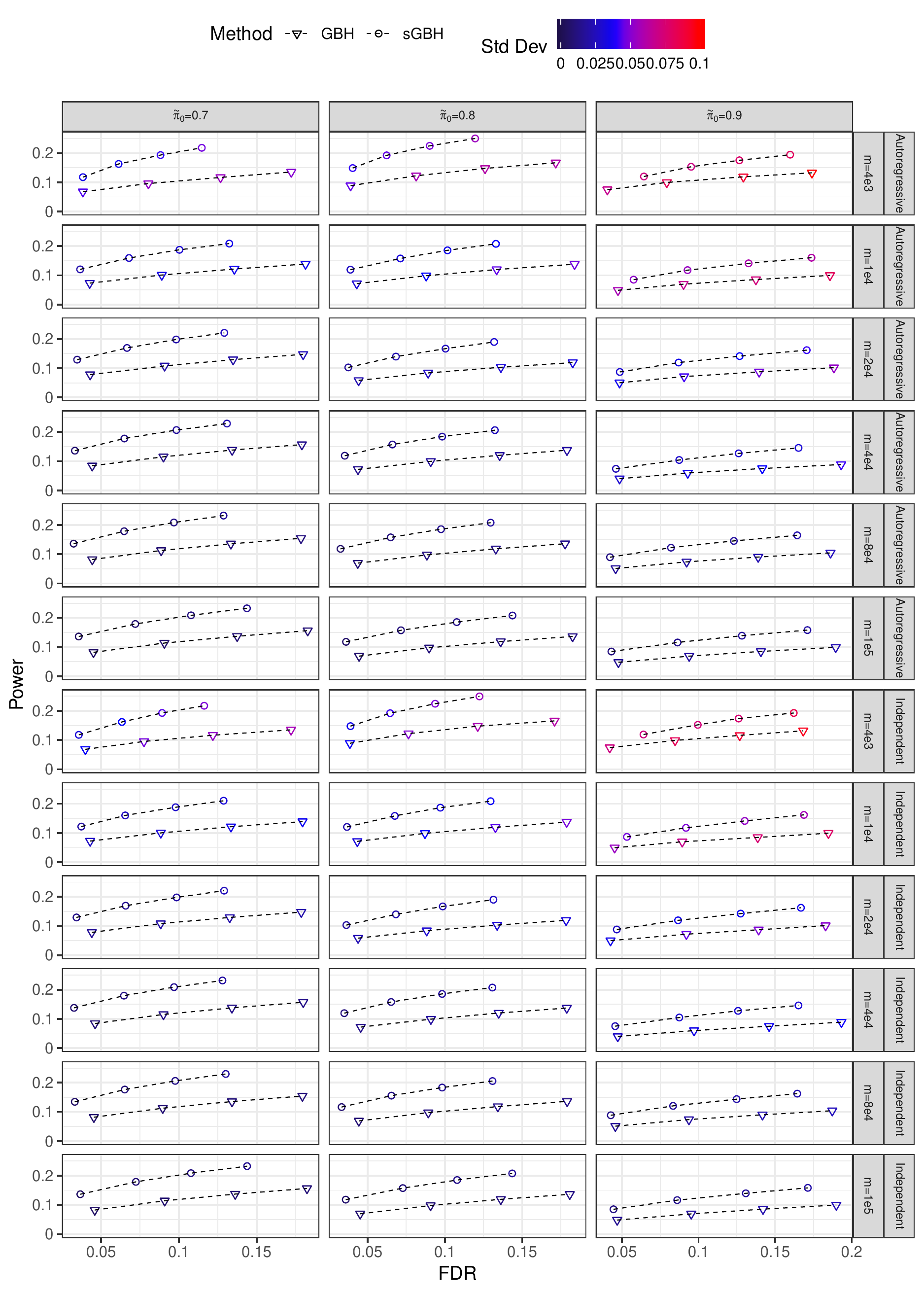}
\vspace{-0.5cm}\caption[power new, 1-sided, simes]{FDRs and powers
of the generic adaptive sGBH (``sGBH'') and GBH (``GBH'') based on one-sided p-values and tuning parameter $\lambda=0.5$. Each type of points from left to
right in each subfigure are obtained successively under nominal FDR level
$0.05,0.1,0.15$ and $0.2$. The color legend ``Std Dev'' is the standard deviation of the FDP. Simes test has been used to identify interesting groups at Type I error level $\xi=0.1$.}%
\label{figPowerNewSimes1side}%
\end{figure}

\begin{figure}[H]
\centering
\includegraphics[height=0.87\textheight,width=.95\textwidth]{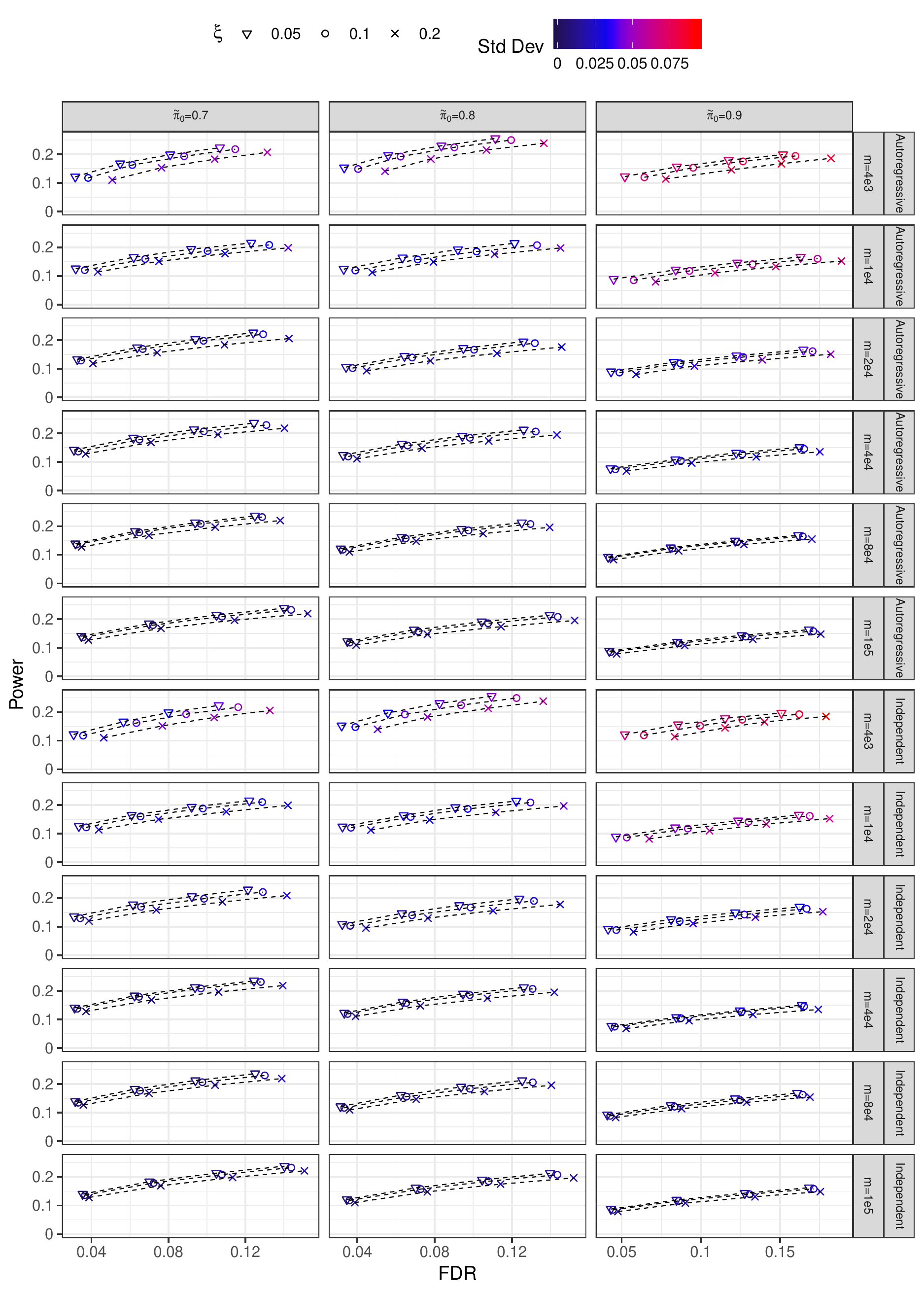}
\vspace{-0.5cm}\caption[power new, 1-sided, simes, xi]{FDRs and powers
of the generic adaptive sGBH (``sGBH'') and GBH (``GBH'') based on one-sided p-values and with tuning parameter $\lambda=0.5$ and as the Type I error
level $\xi$ (shown in the legend) of Simes test ranges in $\{0.05,0.1,0.2\}$. Each type of points from left to
right in each subfigure are obtained successively under nominal FDR level
$0.05,0.1,0.15$ and $0.2$. The color legend ``Std Dev'' is the standard deviation of the FDP. Simes test has been used to identify interesting groups.}%
\label{figPowerNewSimesXi1side}%
\end{figure}


\begin{figure}[H]
\centering
\includegraphics[height=0.85\textheight,width=\textwidth]{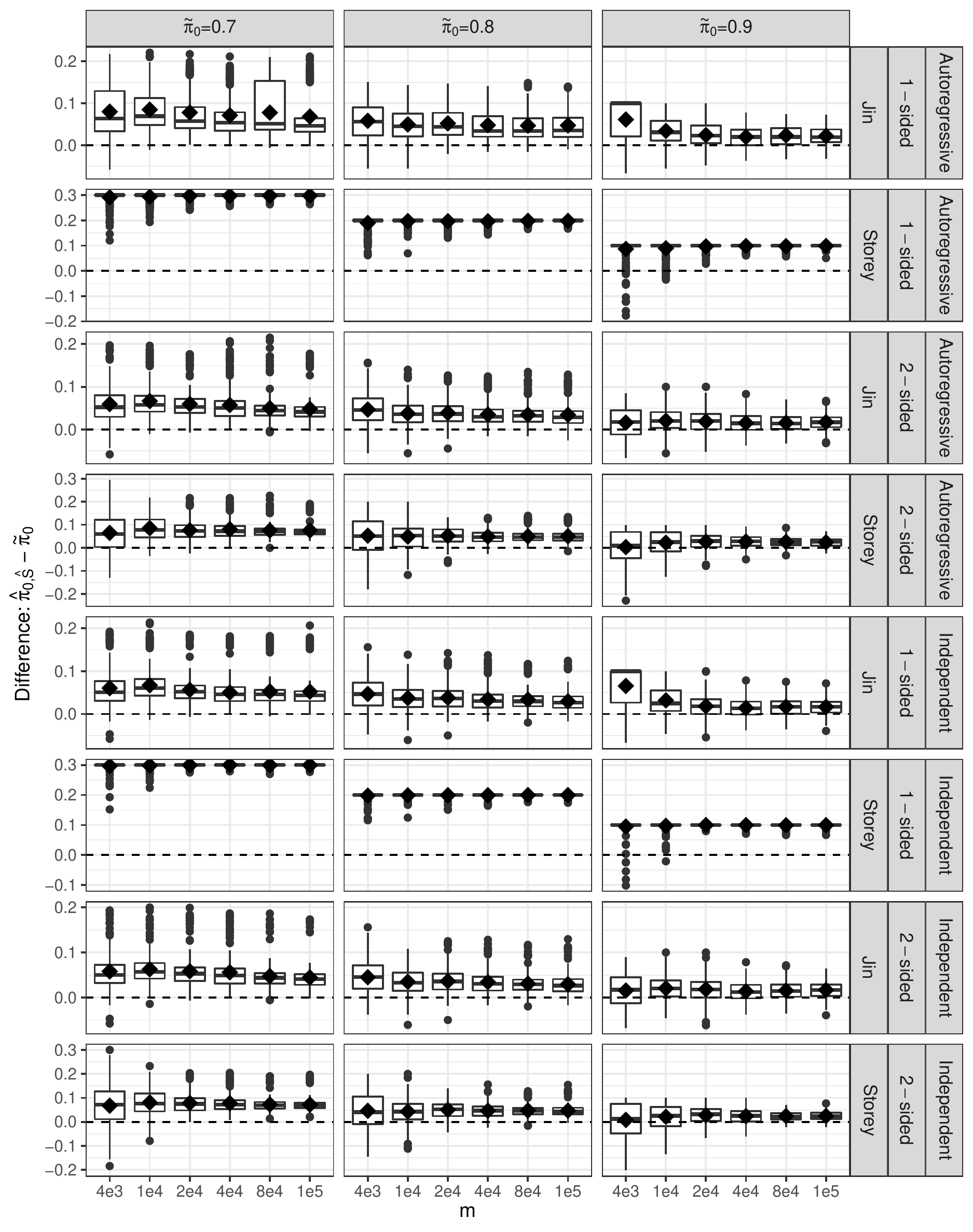}
\vspace{-0.5cm}\caption[pi0Sest]{Boxplots of the estimated proportion for the
interesting group. The estimators ``Jin" and ``Storey" in the strip refer to Jin's estimator and Storey's estimator. In each
boxplot, the diamond indicates the mean of the corresponding estimate.}%
\label{figGroupPi0EstStorey}%
\end{figure}

\begin{figure}[H]
\centering
\includegraphics[height=0.88\textheight,width=\textwidth]{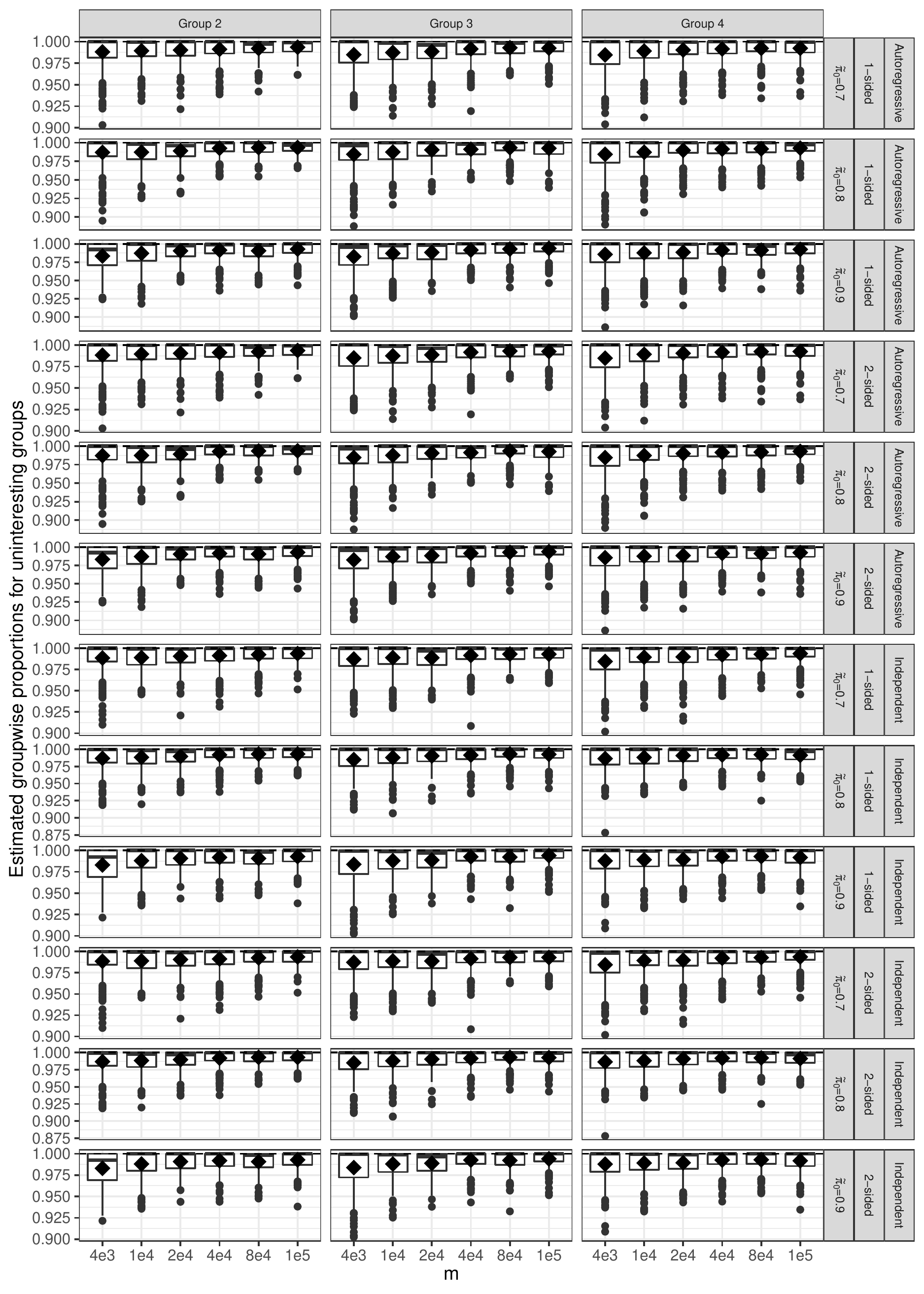}
\vspace{-0.5cm}\caption[pi0est for uninteresting groups, jin's est]{Boxplots
of the estimated groupwise proportions for the three uninteresting groups. The
estimates are given by Jin's estimator. In each boxplot, the diamond indicates
the mean of the corresponding estimate.}%
\label{figGroupPi0EstInd}%
\end{figure}

\begin{figure}[H]
\centering
\includegraphics[height=0.88\textheight,width=\textwidth]{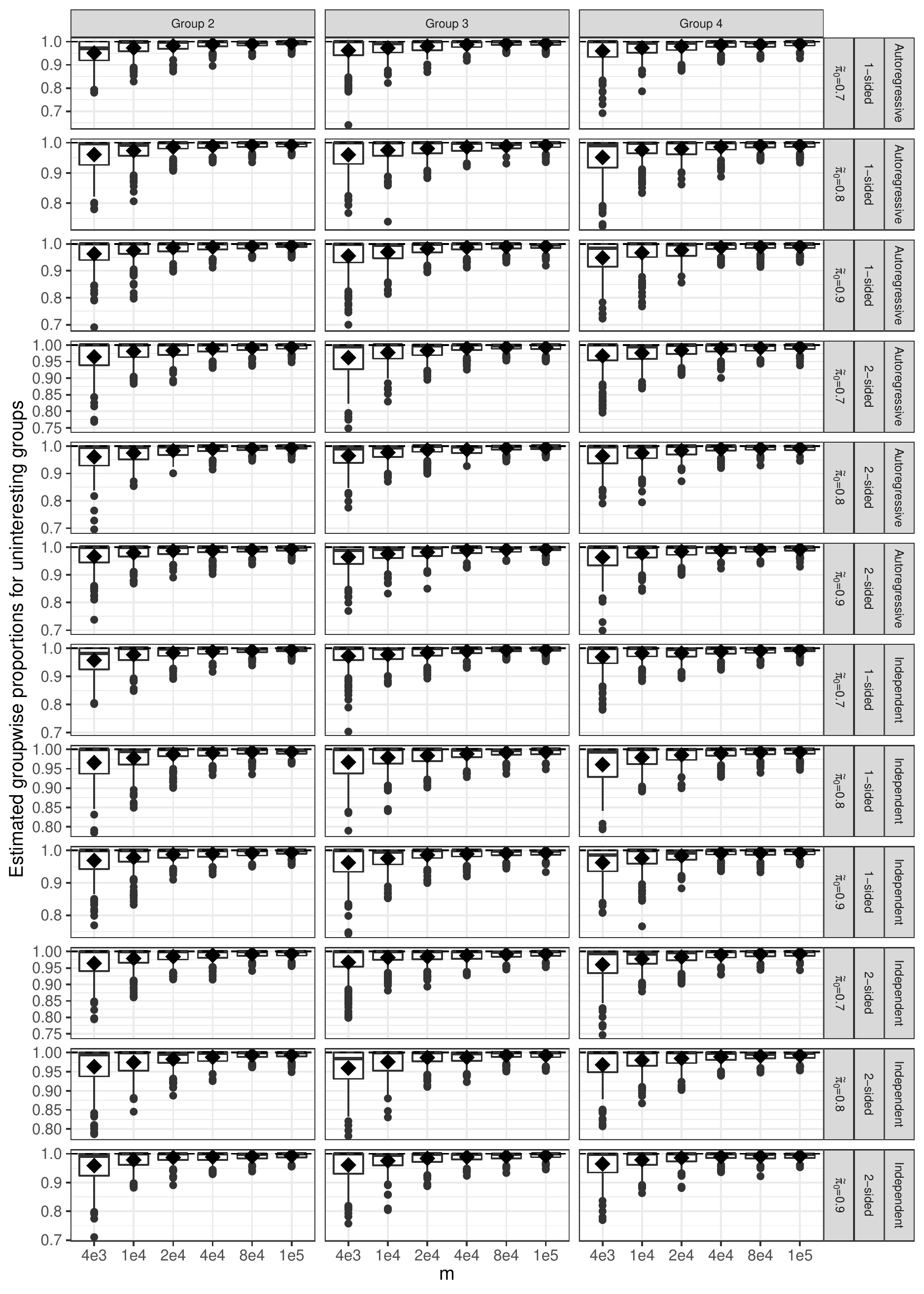}
\vspace{-0.5cm}\caption[pi0est for uninteresting groups, storey's
est]{Boxplots of the estimated groupwise proportions for the three
uninteresting groups. The estimates are given by Storey's estimator. In each
boxplot, the diamond indicates the mean of the corresponding estimate.}%
\label{figGroupPi0SEst}%
\end{figure}

\end{document}